# Resolving the pressure induced 'self-insertion' in skutterudite CoSb$_3$

Bihan Wang[1,2], Anna Pakhomova[1,3], Saiana Khandarkhaeva[4], Mirtha Pillaca[5], Peter Gille[5], Zhe Ren[1], Dmitry Lapkin[1,6], Dameli Assalauova[1], Pavel Alexeev[1], Ilya Sergeev[1], Satishkumar Kulkarni[1], Tsu-Chien Weng[2], Michael Sprung[1], Hanns-Peter Liermann[1], Ivan A. Vartanyants[1], Konstantin Glazyrin[1,*]

[1]Photon Science, Deutsches Elektronen-Synchrotron DESY, Notkestr. 85, 22607 Hamburg, Germany

[2]Center for Transformative Science, Shanghai Technical University, Shanghai, China;

[3]European Synchrotron Radiation Facility, avenue des Martyrs 71, 38043 Grenoble, France

[4]Bayerisches Geoinstitut, University of Bayreuth, Universitätstr. 30, 95440 Bayreuth, Germany

[5]Ludwig-Maximilians-Universität München, Department of Earth and Environmental Sciences, Crystallography Section, Luisenstr. 37, 80333 München, Germany

[6]Current address: Institute of Applied Physics, University of Tübingen, Auf der Morgenstelle 10, 72076 Tübingen, Germany

Key words: skutterudites; high pressure; diamond anvil cell; single crystal diffraction; Bragg coherent diffractive imaging

**Abstract**

CoSb$_3$ belongs to the skutterudite family of compounds and serves as a crucial platform for the exploration of thermoelectric materials. Under compression it undergoes a 'self-insertion' isostructural transition resulting in a peculiar redistribution of large Sb atoms between different crystallographic sites. We conducted a comprehensive investigation of CoSb$_3$ structural phase stability up to 70 GPa using single crystal material in the regimes of conventional single crystal X-ray diffraction and the X-ray scattering focused on measuring Bragg peak at high resolution (including elements of Bragg Coherent Diffraction Imaging). We explore the compression behavior of CoSb$_3$ in three different pressure transmitting media (PTM) and address several important topics: influence of various PTMs and nonhydrostatic stresses on the strongly correlated system of CoSb$_3$, including the 'self-insertion' crossover, phase stability of CoSb$_3$, the compound's polymorphism, its crystal chemistry, and its variation under pressure. Among other important observations, we track population of Sb atom within CoSb$_3$ dodecahedral sites on compression, during the process of 'self-insertion', and on decompression. We detect that 'self-insertion' may not only reduce the compressibility, but also make it negative. Finally, but not the least, we report that the 'self-insertion' crossover is an important step preceding a previously unknown phase transformation from a cubic $Im\bar{3}$ CoSb$_3$ into a trigonal $R\bar{3}$ occurring above 40 GPa, and discuss the distinctive behavior of CoSb$_3$ phases and their structural frameworks.

*Corresponding author: Konstantin Glazyrin, email: konstantin.glazyrin@desy.de

## 1. Introduction

At ambient pressure, the compound $CoSb_3$ ($Im\bar{3}$, S.G. 204) is known as an exceptional testbed for materials with enhanced thermoelectric properties, thus, there is no surprise that it has attracted significant amount of attention [1–8]. Its electronic, transport and etc properties have been studied for decades with focus on understanding the material in its pure form and the potential for tailoring, e.g. by means of chemical doping. It was a logical progression to investigate the behavior of $CoSb_3$ utilizing the chemically unaltering pathway of compression. Compression paved a novel way for benchmarking theoretical studies and inspiring new avenues for synthesis and property customization [9–11].

The literature review reveals that, despite significant experimental progress in understanding the behavior of $CoSb_3$ at ambient pressure, studies conducted at high pressure conditions are relatively scarce [9,11–13]. We have to admit that we have only limited information with respect to the fundamental information, such as $CoSb_3$ structure as a function of pressure and temperature. In this single crystal X-ray diffraction work we fill missing information and, in particular, address the notable phenomenon of 'self-insertion' occurring above 25 GPa which is clearly under-characterized even under ambient temperature conditions.

The definition of 'self-insertion' is attributed to a distinctive isostructural phase transition reducing the $CoSb_3$ bulk compressibility, hypothesized being a result of an intricate Sb redistribution between $Im\bar{3}$ Wyckoff *24g* and *2a* sites [9]. It was ascribed to a random reorganization of Sb atoms forming $Im\bar{3}$ $CoSb_6$ corner sharing octahedron network into void dodecahedral *2a* sites of the low-pressure phase. Not only 'self-insertion' changes the material elastic properties, it should affect electronic properties [12]. Considering a broader perspective and a bigger family of skutterudites, the 'self-insertion' is not a unique feature which can be attributed only to $CoSb_3$. Matsui et al. presented evidence for 'self-insertion' for $RhSb_3$ and $IrSb_3$ [13]. Apart from thermoelectric applications [11], the process of 'self-insertion' is a very special case of a highly interesting strongly correlated phenomenon, but there are inconsistencies preventing us from crafting a cohesive picture. For example, previous studies reporting 'self-insertion' (e.g. [9,13]) were done in methanol:ethanol mixture pressure medium (highly non-hydrostatic above 10-12 GPa) and provided only selective and incomplete structural information.

In our pursuit to explore the pressure-induced structural evolution of $CoSb_3$ and to unveil new pieces of a bigger puzzle, we studied the same material in various pressure transmitting media (e.g. He, Ne, Ar, sorted according to the atomic number and radii). This led to enhanced characterization of the material and its sensitivity to undesirable stresses and strains. It is known that phase stability in strongly correlated systems can be very sensitive to energy landscape complicated by experimental stress/strain conditions [14,15]. Within the data reported by Kraemer et al. [9] and Matsui et al [13] (see also [16]) we detected evidence for associated effects in inconsistencies for compressibility and equation of state (EoS) above 10 GPa, and in our study we carefully approach the topics of the crossover starting pressure, its width and correlate our observations with 'self-insertion' at various experimental conditions.

Given the large size of target crystallographic sites for 'self-insertion' (e.g. $Im\bar{3}$ phase dodecahedral sites), we also explored the potential for pressure transmitting media (PTM) to penetrate and occupy the latter, as it could potentially affect the process and the final state. This aspect has not been addressed before, but our analysis of conventional single crystal X-ray diffraction provides a thorough answer to this question.

Using various PTMs offers one approach to exploring the impact of undesirable stresses and strains in realistic experimental loadings, but it represents only one aspect of the overall picture. In our studies we also varied sample size from bulk single crystals to much smaller sub-micron particles prepared by a focused ion beam (FIB) technique.

Our experiments conducted on larger single crystals enabled estimation of material micro and bulk compressibility (e.g. bond distance and lattice contraction), characterization of 'self-insertion' and the resulting Sb occupancy of *2a* sites as a function of compression and decompression. To our knowledge, this is the first of the kind single crystal study covering the pressure range below 70 GPa. Using sub-micron FIB milled crystals allowed us to explore in greater detail the starting pressures of 'self-insertion' and compare obtained results with data on bulk crystals. Indeed, within our 'top-to-bottom' preparation procedure, the resulting sub-micron sample size allowed us to increase the volume ratio between PTM and sample in attempt to reach the best quasi-hydrostatic conditions from experimental point of view. These samples were studied with single crystal X-ray diffraction at large sample to detector distance (high resolution).

At the same time, by taking advantage of working with sub-micron sample, we also explored an option of Bragg Coherent Diffraction Imaging (BCDI) [17–21]. The technique is novel for the high-pressure community working with diamond anvil cells (e.g. significant stresses and strains within samples as well as in between samples and sample environment), but presents potential. Below, we show and discuss with the community an important example and provide new observation and a new reference to the emerging field employing BCDI at extreme conditions.

Finally, in our work we present a refined and extended picture of $CoSb_3$ phase diagram with pressures up to ~70 GPa at ambient temperature. Previous works were done on powder material compressed in diamond anvil cell, thus, our results showing in-detail variation of $CoSb_3$ crystal structure in response to compression provide a reference for future complicated theoretical benchmarks involving atomic occupational disorder. Among other results, we indicate that the process of 'self-insertion' is an intermediate step of compression induced transformation of $CoSb_3$, with a new phase transition $Im\overline{3} \rightarrow R\overline{3}$ starting at ~40 GPa, following a group-subgroup path and resulting in an extended filling of dodecahedral sites in a contrast to partially filling within 'self-insertion' pressure range. Below, we present experimental details and then explore and describe our observations following peculiar evolution of $CoSb_3$ structure and its framework building blocks under pressure.

## 2. Experimental details

$CoSb_3$ material was the same as used in [22,23]. For conventional single crystal X-ray diffraction (CSC), we selected small shards broken off bigger crystals. Typically, the size of the shards did not exceed $30 \times 15 \times 5$ µm$^3$ to ensure the best quality loading in terms of quasihydrostaticity, without sacrificing the signal-to-noise ratio. Hereafter, we will refer to these crystals as bulk crystals or CSC crystals.

Symmetric diamond anvil cells (DAC) equipped with Boehler-Almax diamonds [24] with ~56-58 ° of effective X-ray conical aperture were used for pressure generation. Sample chambers of diamond anvil used in this study were prepared by indenting 250 µm thick Re foils. The thickness of initial indentation and the initial sample chamber hole drilled in the gaskets was varying depending on diamond culet size. For diamonds with culet diameter of 300 µm, we used gasket hole diameters of 150 µm and indentation thickness between 35 and 60 µm while for diamonds with culet diameter of 200 µm these values were 100 µm and 30-36 µm, respectively.

Sub-micron samples for high-resolution single crystal diffraction (HSC) were prepared in a different way. We prepared these samples by using focused ion milling device of DESY NanoLab (Hamburg, Germany) similar to the procedure described in [22]. The preparation started by first cutting a lamella with a size of $14\times10\times0.5$ µm$^3$. Then, we mounted the lamella to a 2 µm thick amorphous silicon substrate (aSiN, produced by Norcada™) and reduced most of its volume, but keeping cube like single crystal particles $0.5\times0.5\times0.5$ µm$^3$ at the corners. A rectangular piece of aSiN holding four $CoSb_3$ sub-micron particles was cut out from a bigger piece of aSiN membrane and then transferred to a diamond culet to its future sample chamber position (additional details, including microphotographs can be found in [16]). Prior to transfer of the samples on a diamond surface, the aSiN support was slightly engraved to enable recognition of sub-micron sample positions with visible light microscopes. Unlike the situation with CSC, where the samples could be easily located by X-ray absorption, in order to enable faster sample finding, a small piece of tungsten was placed as a reference in the vicinity of the aSiN rectangular piece carrying the samples. In the case of HSC we used mxb90 [25] DAC cells with an effective X-ray aperture of ~70°.

As we show below, these small samples, at least at ambient conditions, could also be characterized by BCDI. Still, the primary task for HSC was to create the most quasihydrostatic conditions for a single crystalline grain, ensuring that the amount of pressure medium was substantially larger than the studied object. This is often not the case in most high-pressure studies, including our CSC.

Here, we studied the same material with three different PTM: He, Ne and Ar. Considering CSC, below we report results for loadings with all three PTM. For our HSC loading with sub-micron crystals we used Ne in order to reduce risk of diamond failure. With respect to the latter, Ne is considered as better quasihydrostatic PTM than Ar, but safer for the diamonds than He.

In all cases we used ruby as a pressure marker [26] with a crosschecking by X-ray diffraction from PTM (e.g. Ne, Ar [27,28]).

### 2.1. Conventional single-crystal diffraction

Conventional high-pressure single crystal diffraction was measured at P02.2, PETRA-III, DESY, Hamburg, Germany [29]. Diffraction data were collected at room temperature using the Perkin-Elmer XRD 1621 detector ($200\times200$ µm$^2$ pixel size). The wavelength was changing depending on a dataset (e.g. 0.2907, 0.2952, 0.2955 Å). The single-crystal step-scans were collected during a continuous rotation along a vertical axis ω within ±28-29 °, depending on a DAC X-ray aperture, with a step varying within 0.5-1°, depending on a dataset. We also collected oscillation images covering ±20 ° of rotation within a single frame. Examples of 2D oscillation patterns can be found in Supplementary [16].

The 2D patterns were inspected and analyzed using DIOPTAS software [30]. Indexing and intensity integration was performed using CRYSALIS PRO [31], and for crystal structure solution we employed JANA2006 and OLEX2 [32,33] with SHELX backend [34,35]. Inspection of 1D patterns integrated from 2D images using DIOPTAS was also conducted with JANA2006. During preparation and writing the manuscript we additionally used VESTA [36] and CRYSTALMAKER [37] for the purpose of structure visualization.

The information covering crystal structure solution is presented within the tables of Supplementary [16]. In addition to the tables, in Supplementary we provide Crystallographic Information Files (CIFs). We have to note that the atomic occupational disorder associated with 'self-insertion' was triggering significant diffraction peak profile broadening, thus, after the crossover, the data signal-to-noise ratio was reduced. The effect can be seen within diffraction patterns shown in Supplementary and should not be confused with other potential effects, e.g. crystal bridging between diamond anvils, etc. The reduction of signal-to-noise ratio affected fit quality and resulted in higher values of crystallographic R-factors. Still, we consider that our analysis sufficiently reflects the structural evolution under high pressure conditions, and that the major conclusions remain unaffected. Considering the stability field of $R\bar{3}$ high pressure phase discussed below, due to experiment beamtime limitations we could only measure its crystal structure at 68.8(2) GPa. The data points collected between 41 and 68 GPa were recorded in a form of 2D oscillation patterns providing reduced angular information (e.g. without any correlation between Bragg peak position vs DAC rotation angle). Given a large number of observations, this information was still sufficient to calculate $R\bar{3}$ lattice unit lattice parameters.

### 2.2. High resolution single crystal diffraction and BCDI.

The diffraction patterns from sub-micron particles were recorded at Coherence Application P10 beamline of PETRA-III. We used a coherent 13.1 keV X-ray beam focused to approximately 1×1 µm² using Compound Refractive Lenses (CRL) as focusing optics. Prior to data collection, the sub-micron crystals enclosed within a DAC, were aligned to the rotation center of the diffractometer, with the sample-to-detector distance fixed to 1.83 m, with sample-signal scattering in the horizontal plane. The data of a pre-aligned $CoSb_3$ $Im\bar{3}$ (013) Bragg peak were measured using EIGER X4M detector (75×75 µm² pixel size). The $Im\bar{3}$ (013) peak was considered as the most intense peak of the structure fitting the aperture of a DAC at 13.1 keV at all pressures of the study. In addition, our simulations indicated a strong correlation between $Im\bar{3}$ (013) and the 'self-insertion' crossover making (013) a strong signal source for the crossover starting pressure determination.

At each pressure point, including the ambient, we collected a full snapshot of $Im\bar{3}$ (013) Bragg peak intensity by step scanning. The geometry of scattering in horizontal plane is depicted in **Figure 1**. At each rotation step Δθ, the detector captured a 2D diffraction pattern. These patterns were later assembled into a 3D diffraction pattern dataset corresponding to the intensity distribution in reciprocal space. A typical collection involved collection of ~300 2D diffraction patterns within ±1 ° scan range.

Ambient conditions real-space images were reconstructed from the 3D intensity distribution at ambient pressure (see discussion below). The pyCXIM package [38] was used to process the iterative algorithm which consisted of a sequence of ER350 + HIO800 + RAAR2100 + DIF1100 (where the following abbreviations were used: Error-Reduction (ER), Hybrid Input-Output (HIO), Averaged alternating Reflectors (RAAR), Difference Fourier (DIF)). At every iteration, modulus and support constrains were applied. The shrink-wrap algorithm [39] was also included with the threshold range used in analysis that was in the range of 0.08-0.11. The best reconstructions were considered to be those with minimum Fourier space error.

## 3. Results and Discussion

### 3.1. Polymorphism of CoSb$_3$ below 70 GPa and sensitivity of phase transitions to sample environment

By means of single crystal X-ray diffraction we could track diffraction signal and correlated parameters with fine accuracy and precision. Considering initial $Im\bar{3}$ phase, we report that the Bragg peak width of (013), attributed to $Im\bar{3}$, is very sensitive to changes of thermodynamic conditions. In **Figure 2** we show compression induced change of the (013) Bragg full width at a half maximum (FWHM) as a function of pressure for Ne pressure medium for bulk and sub-micron crystals. We note that although for the case of CSC we used all equivalent reflections contributing to our oscillation images, for sub-micron crystals we had access to a single one of them.

It is clear that for all four sub-micron crystals loaded into a DAC, the $Im\bar{3}$ (013) FWHM increases with a pressure change from 24.3(2) to 29.7(2) GPa. We attribute it to the start of the 'self-insertion' for this HSC loading and provide an additional illustration in **Figure 3**. The significant broadening of the (013) peak profile as projected to the d-spacing axis at pressures above 25 GPa cannot be simply explained by an average particle strain (bulk strain), as asymmetry of the Bragg peak changes. Indeed, at lower pressures, we observe a small tail at higher d-spacing values, while at higher pressures we see a stronger broadening and asymmetry developing at the opposite side.

Our HSC data are limited by ~30 GPa, but within CSC data we can find additional evidence. FWHM of the (013) reflection in a bulk crystal (CSC) loaded with Ne begins to broaden at pressure of 32.4(2) GPa, thus, at higher pressure in comparison to HSC loaded with same PTM. As indicated using shading in **Figure 2**b, the broadening of CSC reaches saturation below ~40.9(2) GPa. Further increase of the (013) FWHM is observed at higher pressures.

Considering the pressure range below 70 GPa, in contrast to Matsui et al. [13] (powder diffraction data, methanol-ethanol mixture PTM), instead of two distinct pressure regions we highlight three: (1) below the 'self-insertion' when Sb population of $Im\bar{3}$ $2a$-site is 0, (2) during the 'self-insertion' when Sb occupancy of $2a$-sites becomes non-zero and (3) when CoSb$_3$ $Im\bar{3}$ is transformed via a group-subgroup transition into a lower symmetry structure. The latter conclusion is supported by our observation of $Im\bar{3}$ (013) splitting described in greater detail in Supplementary [16]. Analysis of our CSC data allows us to index the new structure as $R\bar{3}$ (S.G. 148) with Z=12 (formula units per unit cell) in contrast to Z=8 for $Im\bar{3}$.

Before we come to the description of material compressibility and $R\bar{3}$ structure, we clarify several important points. First of all, we note, that our CSC data excludes any incorporation of

He, Ne or Ar at dodecahedral sites of $CoSb_3$ polymorphs as tested in multiple loadings and supported by our crystal structure refinements as well as by observations of material compressibility discussed below.

A few words on the Sb occupancy of dodecahedral sites triggered by 'self-insertion'. We report that the measured values correlate with our observations of the (013) broadening at as we indicate in **Figure 4**. Data shown in the figure support our conclusions on 'self-insertion' starting pressures. For the simplicity of the representation we only show the experimental points close to the 'self-insertion' and skip lower pressure points preceding the crossover. For the sake of completeness, we note that in order to fully fill the dodecahedral sites of a $Im\bar{3}$ unit cell we require only two Sb vacancies at $Im\bar{3}$ *24g* sites of Sb per unit cell (Sb forming octahedral arrangement around Co). The final Sb occupancy of dodecahedral sites, prior to a full transition to $R\bar{3}$ does not reach 100 %, indicating that the degree of 'self-insertion' is an important factor for the $Im\bar{3} \rightarrow R\bar{3}$ transition. Finally, we note that decompression to ambient of a CSC sample loaded in Ar pressure medium shows a clear hysteresis with final antimony occupancy at dodecahedral sites reaching a value of 28(1) %.

Next, if we compare our data with literature, we see that the starting pressures of 'self-insertion' of HSC compressed in Ne and CSC compressed in Ar and Ne pressure media differ from one another significantly. In all our loadings 'self-insertion' starts at values reproducibly higher than the values reported in the original publications of Kraemer et al. [9] and Matsui et al. [13]. It is also worth to note that in contrast to Kraemer et al. [9] and Matsui et al. [13] (powder material, methanol-ethanol mixture PTM), our data reproducibly indicates that prior to $Im\bar{3} \rightarrow R\bar{3}$ transition, the 'self-insertion' transition occurs over a relatively narrow pressure range.

In this work, and for the first time, we present data on a $CoSb_3$ with an outlook to the details of sample environment (different pressure media and different stress/strain relations for single crystalline samples). Our data on sub-micron and bulk crystalline material in comparison to literature indicates extreme sensitivity of $CoSb_3$ to external stresses and surrounding high-pressure environment which should also be carefully considered for other skutterudite materials, including $RhSb_3$ and $IrSb_3$ reported by Matsui et al. [13] and etc.

In most of conventional high-pressure experiments, the important details of realistic sample loadings with their stresses and strains are seldom discussed. As we demonstrate here, in strongly correlated systems like $CoSb_3$, these factors play a significant role and should not be neglected. This is important not only for theoretical studies but also for benchmarking experiments and vice versa.

### 3.2. X-ray diffraction with coherent beam and high resolution: investigation of sub-micron CoSb₃ particle at ambient and elevated pressures

CSC under extreme conditions has advantages and limitations. In comparison to HSC, it has much lower detection limit of initial sample strain conditions. Investigations of X-ray diffraction under high resolution and application of BCDI can help us to shed light on initial micro-strain conditions of the sub-micron samples. We consider important sharing these observations with the community because of several reasons. On one hand we want to attract community attention to the power and limitations of BCDI applied at high pressure. On the other hand, it is also one of the rare cases when an extreme conditions sample analyzed by BCDI was prepared using a 'top→bottom' approach in which a smaller particle was cut from a much larger crystal.

First of all, in **Figure 5** we show data for a submicron particle transferred together with as aSiN support into a sample chamber of a DAC. The data were measured at ambient pressure and temperature prior to a loading with PTM. The HSC 3D sub-micron particle shape and the strain component $\varepsilon_{zz}$ with direction of Z along the scattering vector (013) of $Im\bar{3}$ were determined by phase retrieval from the 3D diffraction pattern. The voxel size of our reconstruction given by the dimensions of probed reciprocal space was about 16 nm. The data shown in **Figure 5** represents a sum of 50 best results out of the 100 reconstructions. The reconstructed nanoparticle can be described as a cuboid with rounded corners and edges with a characteristic edges side of ~ 540 nm. This value is in a good agreement with measurements conducted with the FIB dual beam machine. Our analysis resolves some strain at the surface of the nanoparticle potentially induced by the preparation procedure of FIB cutting. Still, these values are minimal in comparison to the more homogeneous central part. Based on our observations of a case using 'top→bottom' sample preparation approach for BCDI, we suggest that the concepts of sample pre-characterization, sample redundancy and experimental reproducibility become critical. Pre-characterization of multiple BCDI samples requires significant amount of beamtime which should be carefully considered by high-pressure community.

Considering compression of the same sample in Ne, the stress induced changes of diffraction signal are well seen in **Figure 6**. After comparing comparison of intensity distribution corresponding to different pressure points, we detect significant differences between the recorded signals. We observe the effect of compression starting already from 1.6(2) GPa. The signal becomes more and more complicated with each pressure step. The strong contrast change between the points of 24.3(2) GPa and 29.7(2) GPa contributes to our discussion of developing 'self-insertion'. Notably, similar changes are observed for other measured particles mounted to the same aSiN substrate, as demonstrated in Supplementary [16]. Unfortunately, the BCDI reconstruction of high-pressure data did not yield a reliable result for neither of the prepared 'sub-micron' particles subjected to high-pressure conditions in Ne PTM.

### 3.3. Compressibility and crystal chemistry and of $CoSb_3$ below 70 GPa

Evolution of $CoSb_3$ unit cell volume normalized to formula units (Z) is shown in **Figure 7**. In previous publications of Kraemer et al. [9] and Matsui et al [13], it has been suggested that the process of 'self-insertion' should be attributed with a rare phenomenon of reduced compressibility. Our data on two independent Ne loading (e.g. orange and green symbols) shows that the observation of negative compressibility, namely with volume of a material unit cell increasing under compression, is also possible (see the pressure region between 30-35 GPa). The contrast between various observations, including the example of loading Ar PTM, requires some additional discussion.

In our work we describe multiple loadings of different sample size loaded into different pressure media. In **Figure 7**, for pressures below the 'self-insertion' crossover indicated using orange shading we see a very good agreement between CSC and HSC data. Although the scatter of data points between different quasihydrostatic loadings is minimal, we still observe a small mismatch of volume-pressure data. For example, it is seen for two different Ne loadings at presumably similar degree of quasihydrostaticity and with sample of similar size (e.g. orange and green triangles). The contrast of the Ne PTM data with data corresponding to Ar PTM loading is stronger. In order to explain different observations, we follow the discussion of stress components distribution complexity in a realistic DAC experiment present even for samples with small sample size [14]. In our study we see that although having single crystalline material may reduce effects of undesirable inter-grain strains, there are situations and systems particularly sensitive to non-hydrostatic stresses at quasihydrostatic conditions (e.g. deviatoric stress, pressure gradients and systematic shift of a pressure marker reading with respect to the stresses at a sample). Our study shows that such effects could play a significant role also in situations when the studied crystals are not bridged by diamond anvils. The factors mentioned here should be given due attention, and the case of the $CoSb_3$ shows that the effects can be very strong. Here, we note that although, the degree of 'self-insertion' completeness is very close for the Ar and Ne loadings (see variation of occupancies as a function of pressure shown in **Figure 4**) the recorded unit cell volumes corresponding to the 'self-insertion' differ significantly. Considering the data collected above 40 GPa, we understand that they could bear even stronger systematic shift due to some presence of undesirable stresses. At the same time, it is important to show these results representing an important example and a new observation reference.

The atomic occupational disorder attributed to 'self-insertion' crossover has been lacking a thorough characterization if we consider crystal chemistry. The body-centered structure of low-pressure cubic $CoSb_3$ phase has only two independent/general atomic coordinates which could be adjusted as a function of pressure or temperature [40]. Both of them are attributed to Sb1 *24g* Wyckoff site with coordinates (0 $Y_{Sb1}$ $Z_{Sb1}$). Considering evolution of $Y_{Sb}$ and $Z_{Sb}$ **(Figure 8)**, we detect a slight change at the beginning of the 'self-insertion' (e.g. Ne and Ar PTM). However, due to the unique stress conditions of the corresponding crystals, we observe a stronger contrast between different PTM loadings at higher pressure. Again, it could be attributed to the fact that Ar is considered to be a less efficient quasihydrostatic pressure medium in comparison to Ne [41], but coupled with enhanced susceptibility of $CoSb_3$ to external disturbances in the region of 'self-insertion'. We see that for compression below ~35 GPa, the data on Ar fits well with the data of Ne. If we compare **Figure 4** with **Figure 8**, we see that 'self-insertion' is attributed to larger unit cell volume values for sample loaded with Ar in comparison to samples loaded with Ne. It should explain the difference between the atomic coordinate values measured for different pressure media within the crossover region, as well as the data on decompression collected from sample subjected to Ar. The evidences presented so far indicate that 'self-insertion' is very sensitive to the surrounding environment and complexity of stress and strain distribution within. We suggest that this conclusion should be generalized for a wide range of unfilled skutterudite systems (e.g. with the unfilled *2a* site $Im\bar{3}$), including the above mentioned $RhSb_3$ and $IrSb_3$. Potentially it could also be applied to partially filled skutterudite systems.

The CSC data collected for $CoSb_3$ allows us to see variation of atomic coordinates as well as Co-Sb and Sb-Sb distances under pressure providing more insight and more information for future theoretical calculation which could in-detail explain strong correlations on the level of electronic interactions and bonding. The corresponding values are shown in **Figure 9**. Upon the 'self-insertion' we observe stiffening of Co-Sb distances forming $CoSb_6$ octahedra for Ar loading and even their expansion for Ne loading, respectively. Although, within $Im\bar{3}$ phase all six Co-Sb distances attributed to CoSb$_6$ octahedron are the same, we observe that the within $R\bar{3}$ their diversity increases, and that the resulting octahedra become more distorted in comparison to the cubic phase.

In comparison to Co-Sb, the character of the shortest Sb-Sb distances does not change significantly with an exception of the 'self-insertion' crossover region, where they seem to stiffen. At pressures above the crossover they continue to decrease monotonously. The lowest values correspond to the Sb-Sb bonds contributing to the shell of dodecahedral sites, while the second longer distances – to the distances interconnecting them (see also Supplementary [16]). Considering the comparison with ambient pressure data, the distances shown in **Figure 9**b are very close to the value of 2.908(1) Å Sb-Sb bond corresponding to pure antimony [42,43], with one Sb-Sb distance of $Im\bar{3}$ CoSb$_3$ being slightly larger and one slightly smaller. After the review of available literature [44–47], we suggest, that Sb-Sb bond for the pure antimony decreases, but stays close to the values observed in our study. This raises the question regarding the distances between Sb atoms occupying the centers of dodecahedral sites, as a result of 'self-insertion', and their closest neighbors, which are other Sb atoms. The corresponding values are shown in **Figure 10**.

While electronic density of a large Sb atoms is highly compressible, we have no doubts that the distances shown in **Figure 10** exceed values attributed to Sb-Sb bond in pure antimony under pressure [46,47]. It is clear that the process of 'self-insertion' forces a very rapid and coordinated isostructural atomic reconstruction over the bulk of the CoSb$_3$ $Im\bar{3}$ material moving the Sb atoms over a distance of ~3.1 Å which is very significant if we consider atomic scale. To illustrate the magnitude, we note that the shortest distance between Na and Cl in rock salt is 2.8169(4) Å at ambient (e.g. an example of a system with large ionic radii [48]). Typical pressure induced reconstructions (e.g. of martensitic origin) are typically much smaller.

As an element of the periodic table, Sb can be classified as a metalloid and unlike smaller pnictogens, it may not always require a strong covalent bond. Following the ab-initio analysis shown within the supplementary to [12], where the authors simulated the intrinsic disorder of 'self-insertion' by introducing synthetic SbCo$_4$Sb$_{12}$ with the additional Sb occupying the dodecahedral site, we could support the hypothesis suggesting that the compound produced by 'self-insertion' undergoes a change of the initial material electronic state. Data shown in **Figure 9** and **Figure 10** indicate that at 70 GPa the distance between the dodecahedron central Sb atoms is still ~0.2 Å larger than the largest Sb-Sb bond distance for Sb atoms forming the dodecahedron shells and interconnections in-between them. Moreover, it is likely that the valence state of Sb filling the dodecahedral sites as a result of 'self-insertion' is close to zero. We also suggest that the binding of Sb occupying the dodecahedral sites with its closest neighbors should be non-covalent even at pressures as high as 70 GPa, namely within the $R\bar{3}$ phase stability field.

In order to fill all dodecahedral sites of $Im\bar{3}$ phase, one requires only two Sb vacancies per a unit cell with the vacancies formed on the cost of Sb surrounding Co. According to our observations, in the vicinity of 'self-insertion' crossover, the diffusion triggered by the process covers large atomic distances, and it happens fast – on the order of seconds or much less, depending on pressure step. We observe that collective 'mass' transport correlated with random redistribution of Sb atom which is in-turn driven by 'self-insertion' occurs under conditions where ambient temperature kinetics are strongly suppressed. This indicates a unique underlying nature of the process.

We suggest that for a more accurate system characterization we need to extend theoretical picture of CoSb$_3$. It is known that at ambient conditions the material should be a weak direct band semiconductor with $Im\bar{3}$ structure [49]. Under compression, but prior to 'self-insertion' the CoSb$_3$ band gap may become indirect and an experiment shows that resistivity increases with compression [50]. We have to admit that the picture is complex, and it is far from being complete. Thorough theoretical investigations of the material with 'self-insertion' induced atomic disorder of Sb are still missing, but they may generalize the picture of strongly correlated CoSb$_3$ and the phenomenon of 'self-insertion' in comparison with other skutterudite materials. Formation of $R\bar{3}$ phase inheriting Sb atomic disorder adds another layer of complexity and also should also be addressed by theory.

From the crystallography point of view, the $Im\bar{3} \rightarrow R\bar{3}$ phase transition follows a group-subgroup path, and in **Figure 11** we show relation between both structures for 40.8(2) GPa which should correspond to the end of 'self-insertion' crossover. A diagram capturing the relation can be found in Supplementary [16]. In **Figure 11**, we indicate occupational disorder of different atomic Sb positions and highlight the occupied dodecahedral sites using the black color. The unit cell of $R\bar{3}$ is larger (Z=12) in comparison to $Im\bar{3}$ (Z=8). The compression induced symmetry step-down results in a formation of additional independent Co and Sb sites. The $R\bar{3}$ Wykhoff symbols for CoSb$_6$ octahedra forming Co are *3b* and *9d*, and for Sb they are *18f*. The Sb atoms occupying dodecahedral sites receive the Wyckoff symbol of *3a*.

As one can see from **Figure 4** and **Figure 11**, at ~40.8(2) GPa the Sb occupancies of dodecahedral sites are slightly below 100 %. We also observe that the corresponding Sb occupancies of CoSb$_6$ octahedra are high, attributed with values of 0.94 or 94 %. As mentioned above, due to the limitations of experimental beamtime we could only measure reliably $R\bar{3}$ crystal structure for the point close to 68.8(2) GPa. Although the resulting *R1* factor is not as low as for lower pressures (e.g. enlisted in Supplementary [16]), indeed the fact of atomic disorder caused diffraction peak broadening and decreased of signal-to-noise ratio, we consider the major observations as being valid. In **Figure 12** we compare the crystal structure modification of $R\bar{3}$ by compression showing different structural projections side by side.

Complementary to **Figure 9**, in **Figure 12** we show the shortest Sb-Sb bond distances by means of green cylinders interconnecting green atoms. Although the bonds shrink as a result of compression, these structural elements remain unchanged through the whole compression range supporting prevalent covalent nature of these Sb-Sb bonds. The dodecahedral sites become fully filled by Sb at 68.8(2) GPa, however, it is not the most striking feature. In comparison to 40.8(2), we see a stronger distortion of the light blue octahedrons in comparison to the pink ones. Both type of octahedrons can be described as CoSb$_{6-\Delta}$ with Δ featuring vacancies at Sb sites surrounding Co. The distortion does not result in a strong change of octahedra volume, as we show in Supplementary [16]. Still, it is clear that the crystal fields of Co occupying the pink and the light blue octahedrons is different. This observation is supporting our hypothesis of a compression induced impactful modification of electronic properties. We note that the light blue octahedra interconnect the fully occupied dodecahedral sites, and together they form channels along the *c*-axis of $R\bar{3}$.

As previously mentioned, CoSb$_3$ exhibits behavior that is in stark contrast to many other known strongly correlated materials. In particular, it should become more insulating under stress [50] while many other materials show the inverse behavior [51,52]. It may be that the structural transformation to $R\bar{3}$ and evolution of its crystal chemistry promotes metallization of the compound, however in order to solve this and other challenges behind the physics of CoSb$_3$ (in particular) and of skutterudite materials (in general) crystallographic experimental data are not enough. We require advancements in the field of theoretical calculations, particularly for materials with structural disorder, which remain challenging and complex.

## 4. Conclusions

Our paper addresses one of the most famous unfilled skutterudite material – CoSb$_3$ used as a platform for thermoelectric applications. Strong correlations of CoSb$_3$ (featured in [49]) also make it an excellent physical platform for studies under high pressure as an analogue to chemical doping. Complementary to previous studies, we present a great amount of data collected on single crystalline material with a focus on the particular 'self-insertion' crossover, which is attributed with a reduced, or as we show in our data potentially even negative compressibility. By presenting data collected from small crystallites surrounded by different PTM we discuss various effects, including extreme sensitivity of CoSb$_3$ to external stress and strain conditions. Among other results, we show that the 'self-insertion' crossover may happen at different stress (e.g. HSC vs SCS), but also at different average strain, namely different unit cell volumes (e.g. Ne vs Ar PTM loading). Our current interpretation corrects earlier understanding presented in previous publications, and establishes a new reference for CoSb$_3$ $Im\bar{3}$ pressure-volume relation below the crossover.

The results collected below 70 GPa allow us to reference the 'self-insertion' crossover as an intermediate isostructural transition of CoSb$_3$ $Im\bar{3}$ preceding the transformation into a novel $R\bar{3}$ phase. Although, CoSb$_3$ was compressed to pressures of 40 GPa and slightly above previously [13], the transformation into $R\bar{3}$ has not been found before. This is most probably due to the effects of large inter-grain strain and disorder within studied powder material resulting in the inability to resolve diffraction peak splitting. We can also mention that powder data would not allow reliable indexing and structure solution, which, to the best of our knowledge, is presented for the $R\bar{3}$ phase for the first time.

In our study, we also discuss progress within our understanding of CoSb$_3$ electronic properties and emphasize that there are many missing puzzle pieces. Our experimental results describing intriguing crystal chemistry of CoSb$_3$ and its evolution under pressure present a great benchmark data for future calculations involving atomic disorder. We hope that our work will inspire more precise and accurate theoretical models shedding light on complex electron correlation phenomena natural not only for CoSb$_3$, but for the general family of unfilled skutterudites.

**Acknowledgements**

We acknowledge DESY (Hamburg, Germany), a member of the Helmholtz Association HGF, for the provision of experimental facilities. Parts of this research were carried out at P02.2 beamline of PETRA III and DESY NanoLab. We would like to thank Thomas F. Keller for assistance in using DESY NanoLab FIB machine. The FIB milling capability was provided via a „Bundesministerium für Bildung und Forschung" grant under contract 05K12WC1 and 05K16WC1 in the framework of the Verbundforschung "Einkristallkristallographie bei hohem Druck und variabler Temperatur" (PI.: Prof. Dr. N. Dubrovinskaia, Uni. Bayreuth). B.W. acknowledges funding provided by Office of the China Postdoctoral Council (OCPC) of the Chinese Ministry of Human Resources and Social Security (MoHRSS) within Helmholtz-OCPC program.

**Figures:**

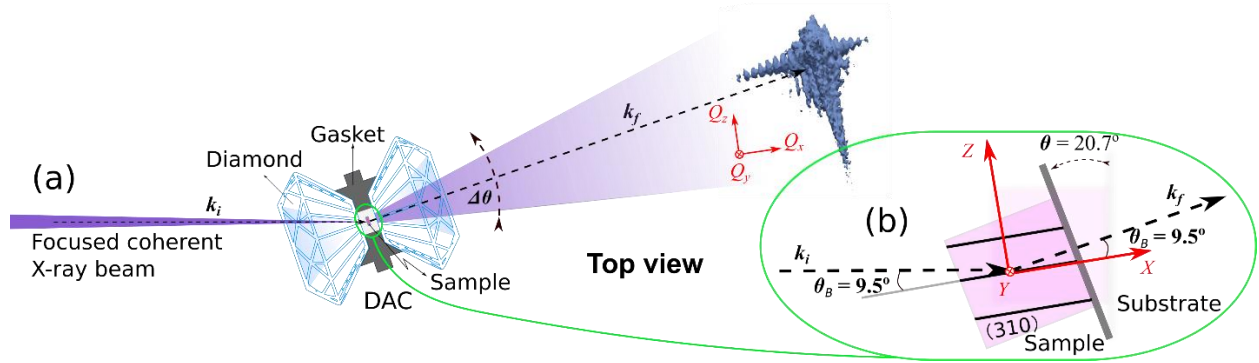

**Figure 1** Single crystal data collection geometry employed for sub-micron samples at the beamline P10, PETRA III. (a) Focused coherent X-ray beam illuminates the sample enclosed within the DAC. Data are collected by rotating the sample centered along θ axis in Δθ steps in order to extract the full $CoSb_3$ $Im\bar{3}$ (013) Bragg peak intensity. In (b) we show a sketch illustrating experimental geometry with the magnified sample mounted to the aSiN substrate.

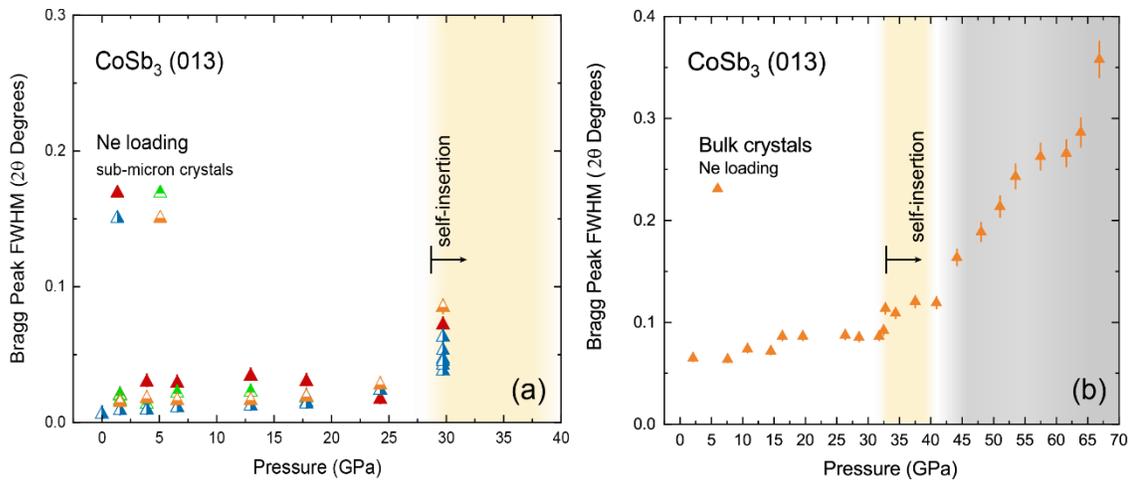

**Figure 2** Selected full width at half maximum (FWHM) data of Bragg peaks attributed to $Im\bar{3}$ CoSb$_3$ (013) at pressures below 40 GPa and its variation under compression. In panel (a) we show the results for sub-micron HSC crystals in Ne PTM collected at 13.1 keV at high resolution, while in (b) we show selected results from bulk CSS loaded using the same transmitting medium and collected at 42.65 keV. Arrows indicate an apparent broadening of the Bragg peak coinciding with the process of 'self-insertion'. Considering four different sub-micron crystals mounted at the same substrate, we see a slight difference in their behavior (e.g. red, orange and blue symbols), but the result is consistent. We note that in (a) the transition happens at earlier pressures in comparison to the data shown in (b). Considering the pressure point of ~29 GPa shown in (a), different blue symbols indicate several different measurements at the same crystal. Finally, for (b) we report that at pressures exceeding 40 GPa, the equivalent reflections (attributed to (013) $Im\bar{3}$ CoSb$_3$ ambient pressure phase) become split indicating a phase transition. We use different shading to separate various pressure ranges. The errorbars are indicated, if not visible, they are the size of symbols or below.

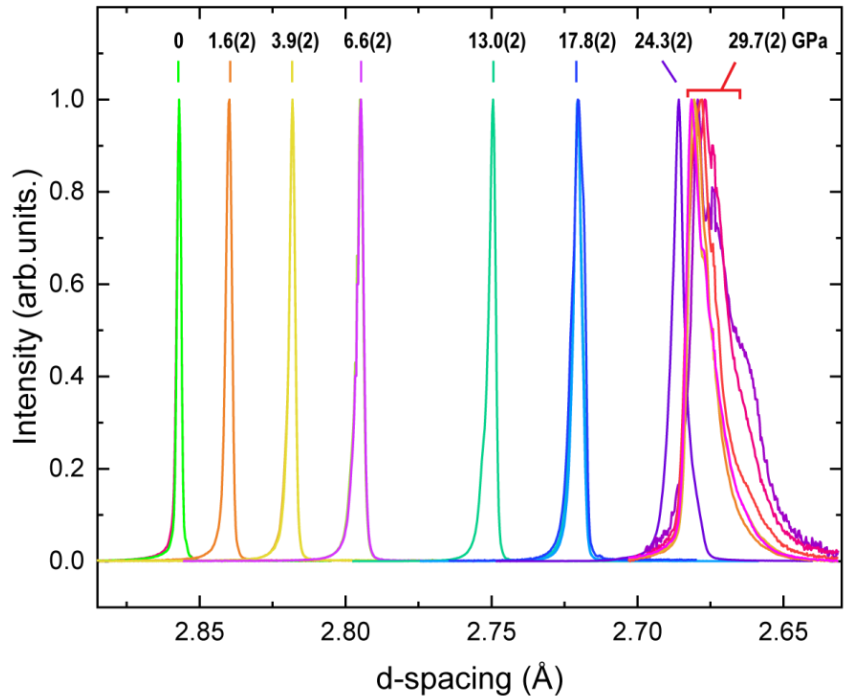

**Figure 3** Variation of CoSb$_3$ $Im\bar{3}$ (013) Bragg peak profile as projected to d-spacing axis. For a simplicity of presentation, we show data from a single sub-micron particle normalized to its maximum intensity. Upon compression, the width of the Bragg peak broadens. Below 24.3(2) GPa, the broadening is visible and could be attributed to the average strain of the particle. For pressure points above 24.3(2) GPa, collected at apparently the same pressure conditions, as indicated by the pressure sensor measured before and after the measurements at P10, we see significant broadening which cannot be exclusively attributed to the average particle stain. Profiles shown for 29.7(2) GPa correspond to several data sets collected on the same particle.

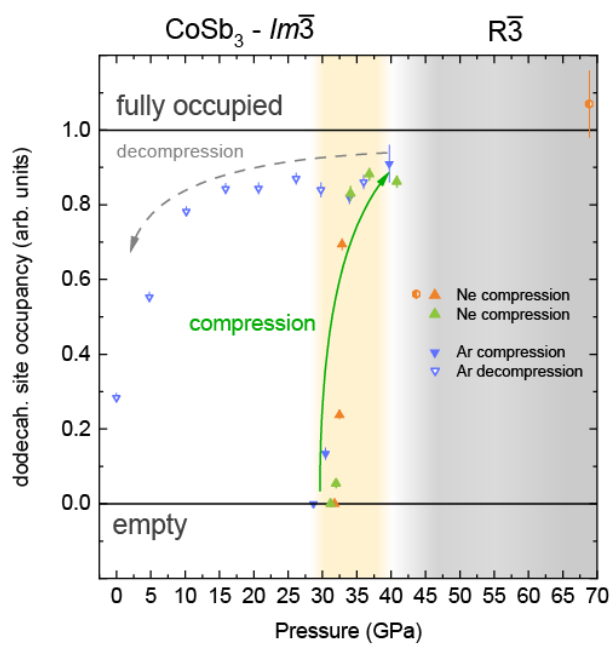

**Figure 4** Antimony occupancies of dodecahedral sites in CoSb$_3$ $Im\bar{3}$ and $R\bar{3}$. Different CSC sample loadings are indicated by symbols with different color. The open blue triangles indicate decompression of CSC in Ar PTM. The orange hexagon corresponds to $R\bar{3}$ structure. In order to show the error bars for the latter, we intentionally allowed the data analysis fitting procedure it to vary above the physically possible value of 1. Grey shading indicates the phase stability field of $R\bar{3}$ phase and yellow shading indicates 'self-insertion' crossover. The occupancy is shown as a dimensionless unit with 0 and 1 corresponding to 0 % and 100 %, respectively. Legend of the figure provides additional information on PTMs used in individual loadings.

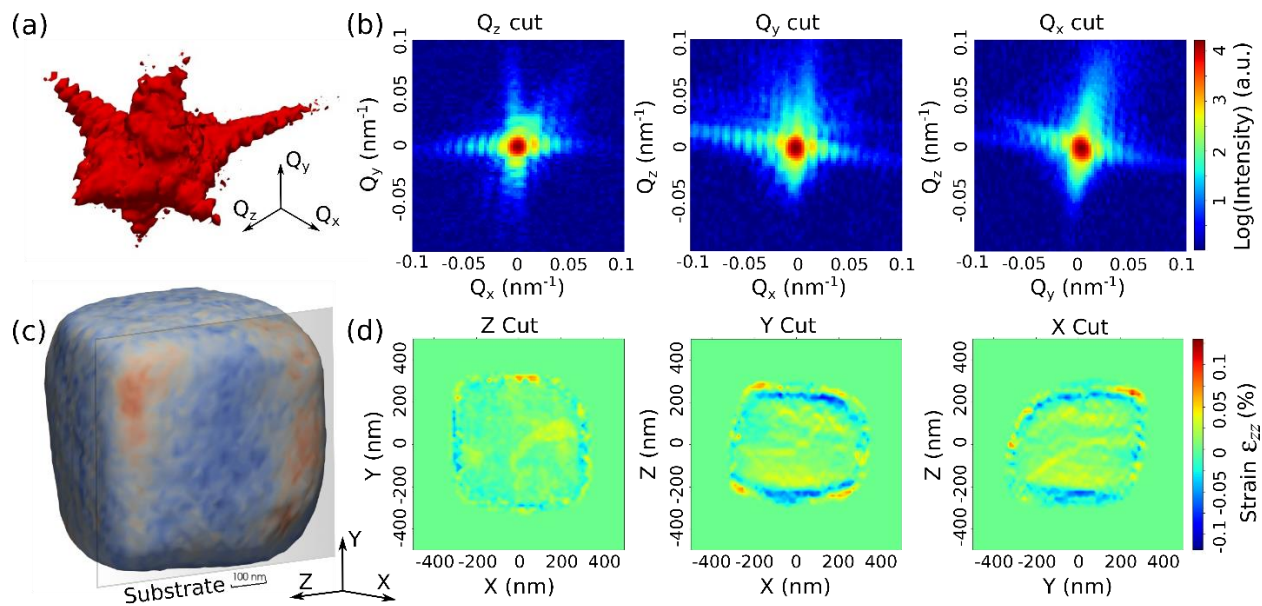

**Figure 5** BCDI analysis of a selected particle loaded in a DAC, mounted to an aSiN substrate at ambient conditions. (a) The 3D representation of the diffraction pattern. (b) The diffraction patterns cut through the center of mass of the intensity in $Q_z$, $Q_y$, $Q_x$. (c) The side view of the reconstructed nanoparticle with surface colored according to the strain field $\varepsilon_{zz}$. The aSiN substrate is shown as a transparent plane with respect to the particle. (d) Illustration of $\varepsilon_{zz}$ strain represented by means of slices going through the center of nanoparticle mass. All diffraction patterns presented in (a, b) are shown on logarithmic scale.

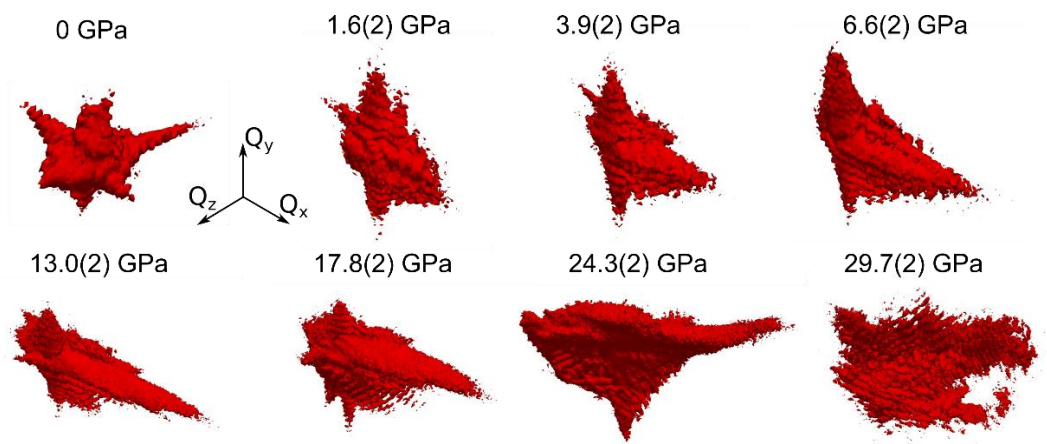

**Figure 6** Evolution of a selected $CoSb_3$ sub-micron particle $Im\bar{3}$ (013) Bragg peak signal as a function of pressure. This representation shares the similar intensity threshold as data shown in **Figure 5**. The particle was compressed in Ne PTM. Even at pressures preceding the 'self-insertion' (e.g. below 29.7(2) GPa), we see a strong effect of compression on the particle signal. While the average strain on the particle is low, the effect of micro-strains accumulated with large pressure steps and at the starting pressure of 'self-insertion' (e.g. 29.7(2) GPa) is clearly visible. The shape becomes much more distorted in comparison to the 24.3(2) GPa point at higher pressure value.

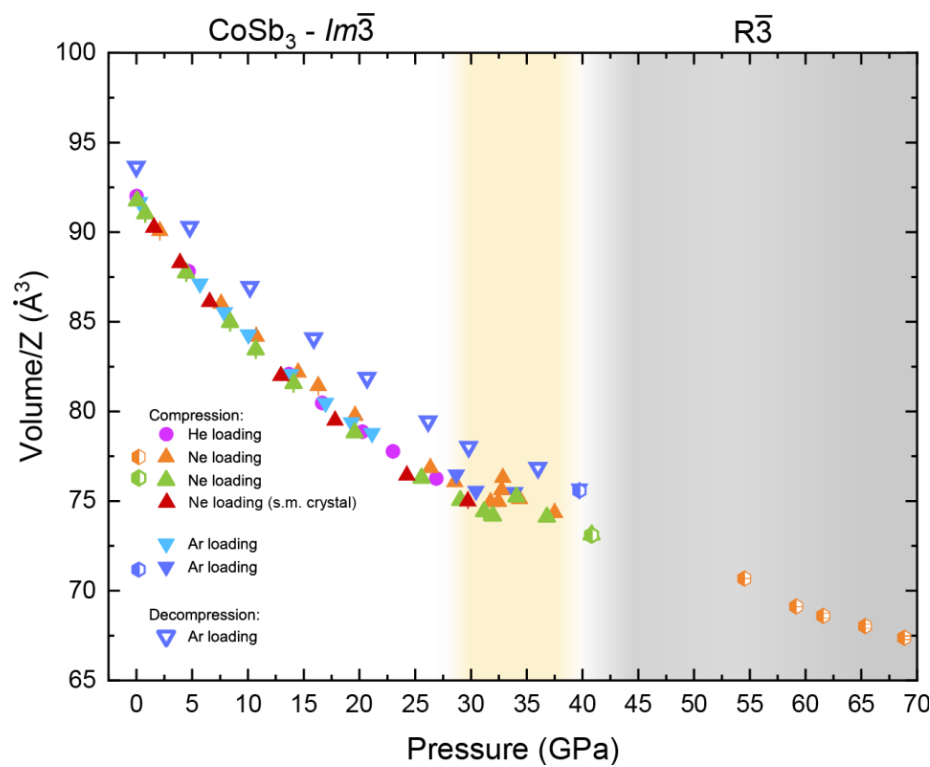

**Figure 7** Compressibility of CoSb$_3$ unit cell normalized to the number of formula units (Z) per unit cell as a function of pressure. Legend of the figure provides additional information on PTMs used in individual loadings. The region of 'self-insertion' highlighted using the orange shading. The phase stability of $R\bar{3}$ is indicated by grey shading. We note good agreement between data from CSC (loaded with various PTMs) and HSC sub-micron crystal. For simplicity of representation we show data for a single particle. Signal for the other HSC particles can be considered as overlaid by the red triangle. Solid circles and triangles correspond to the cubic $Im\bar{3}$ phase. At the same time hexagons correspond to the trigonal $R\bar{3}$ phase. Open inverted triangles correspond to data collected on decompression in Ar PTM, during the same loading as indicated by solid blue triangles. For the reader's convenience, for the selected points at ~40 GPa we present unit cell volumes calculated for both $Im\bar{3}$ and $R\bar{3}$. Our data illustrates that even if the crystals are small and the pressure range is relatively low, there are discrepancies in volume vs pressure for various loadings even if we compare loadings with the same pressure medium. This observation suggests enhanced sensitivity of CoSb$_3$ to external environment. Orange symbols correlate with data shown in **Figure 2** and **Figure 4**. If error bars are invisible, they are the size of the symbols or below.

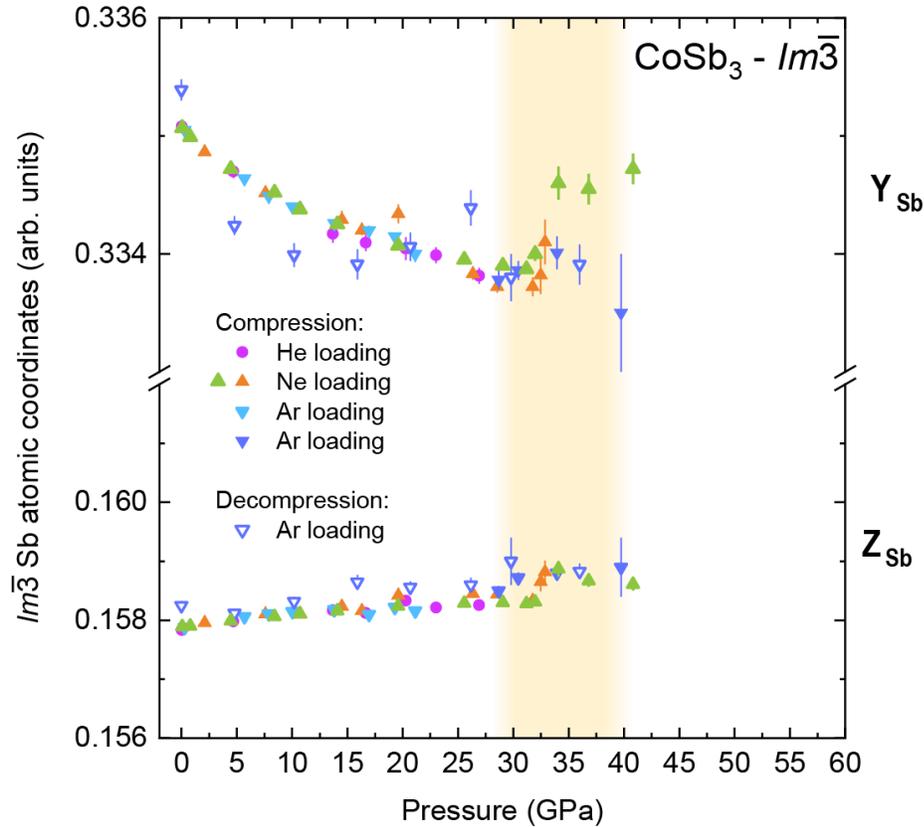

**Figure 8** Evolution of CoSb$_3$ $Im\bar{3}$ Sb atom coordinates as a function of pressure. Solid circles and triangles correspond to the cubic $Im\bar{3}$ phase. Legend of the figure provides additional information on PTMs used in individual loadings. We highlight the region of 'self-insertion' using the yellow shading. Open inverted triangles correspond to data collected on decompression in Ar PTM, i.e. during the same loading as indicated by solid blue triangles. Considering the compression in Ne, we start to see signs for anomalous behavior, namely increasing $Y_{Sb}$ and $Z_{Sb}$ at the early stages of 'self-insertion'. The deviation of data collected in Ar PTM on compression from data collected in Ne PTM could be attributed to the unit lattice mismatch for the given pressure range (see **Figure 7**). Along similar lines, we should attribute the dynamic behavior of $Y_{Sb}$ on decompression to the interplay of various factors, including larger volume of the material undergone 'self-insertion' in comparison to the preceding compression run.

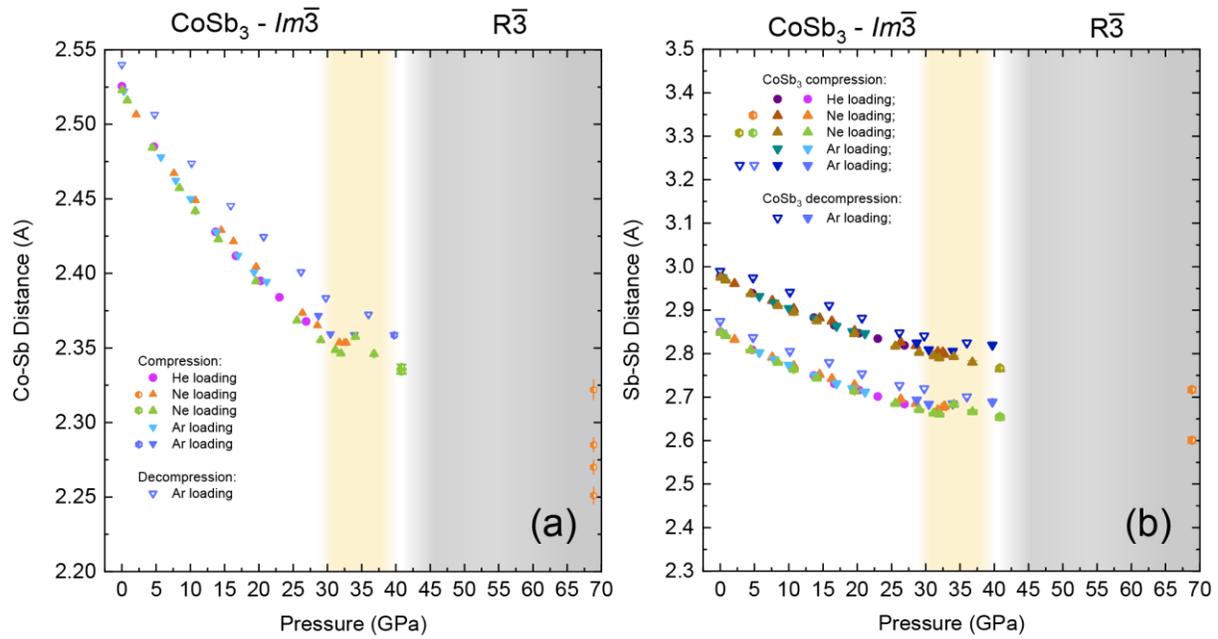

**Figure 9** Variation of shortest near-neighbor (a) Co-Sb and (b) Sb-Sb distances as a function of pressure. We indicate the region of the 'self-insertion' using the yellow shading and the stability field of $R\bar{3}$ using the gray shading, respectfully. The orange hexagons correspond to the trigonal $R\bar{3}$ phase. Open inverted triangles correspond to data collected on decompression in Ar PTM, i.e. during the same loading as indicated by solid blue triangles. For the reader's convenience, for the selected points at ~40 GPa we present values calculated for both $Im\bar{3}$ and $R\bar{3}$. Error bars are shown, if not visible, they are the size of the symbol or below. Legend of the figure provides additional information on PTMs used in individual loadings.

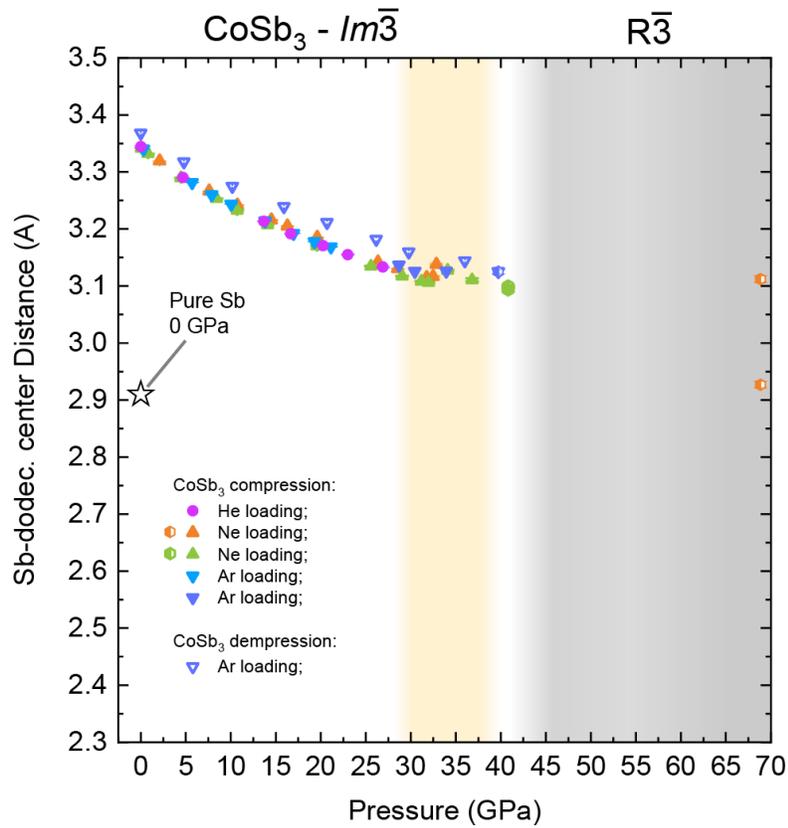

**Figure 10** Distance between a center of a dodecahedral cite (*2a* Wyckoff site of $Im\bar{3}$ or a compatible site in $R\bar{3}$) to the next nearest neighbor represented by other Sb atoms. We indicate the Sb-Sb distance corresponding to pure antimony at ambient conditions using the star symbol [42,43]. We highlight the region of the 'self-insertion' using the yellow shading and the stability field of $R\bar{3}$ using grey shading. Solid circles and triangles correspond to the cubic $Im\bar{3}$ phase. The orange hexagons correspond to the trigonal $R\bar{3}$ phase. For the reader's convenience, we present values calculated for both $Im\bar{3}$ and $R\bar{3}$ for the selected pressure of ~40 GPa. Error bars are shown, if not visible, they are the size of the symbol or below. Legend of the figure provides additional information on PTMs used in individual loadings.

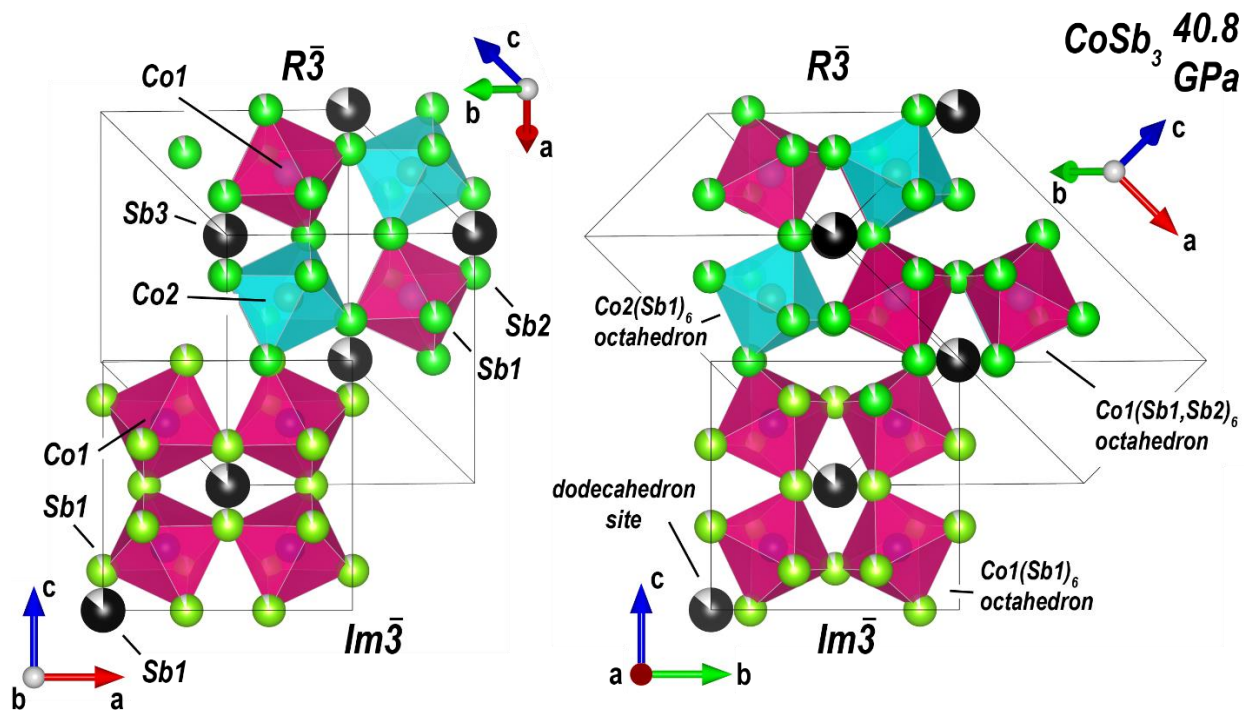

**Figure 11** Comparison between the lower pressure $Im\bar{3}$ and the higher pressure $R\bar{3}$ structures using 40.8(2) GPa as a reference point. The corresponding crystallographic $R1$ factors for $Im\bar{3}$ and $R\bar{3}$ are $R1$=6.64 % and $R1$=6.85 %, respectively. $R$-factors correspond to $I>2\sigma(I)$ with $I$ and $\sigma$ being attributed to the measured intensities and their standard deviations, respectfully. The left and the right panels demonstrate different projections at the same structures. The vectors forming the corresponding unit cells are shown in each panel in the lower left corner for $Im\bar{3}$ and in the upper right corner for $R\bar{3}$, respectfully. The lattice parameters $a$ and $b$ of $R\bar{3}$ lie within a plane perpendicular to [111] of $Im\bar{3}$. By using the black color, we highlight the Sb atoms populating dodecahedral sites and by green – the Sb atoms being part of CoSb$_6$ octahedrons. Occupational disorder of Sb atoms is indicated. Comparing the structures, we observe an addition of independent atomic positions of Co and Sb forming CoSb$_6$ octahedrons in $R\bar{3}$ (e.g. the octahedra painted in light blue and pink).

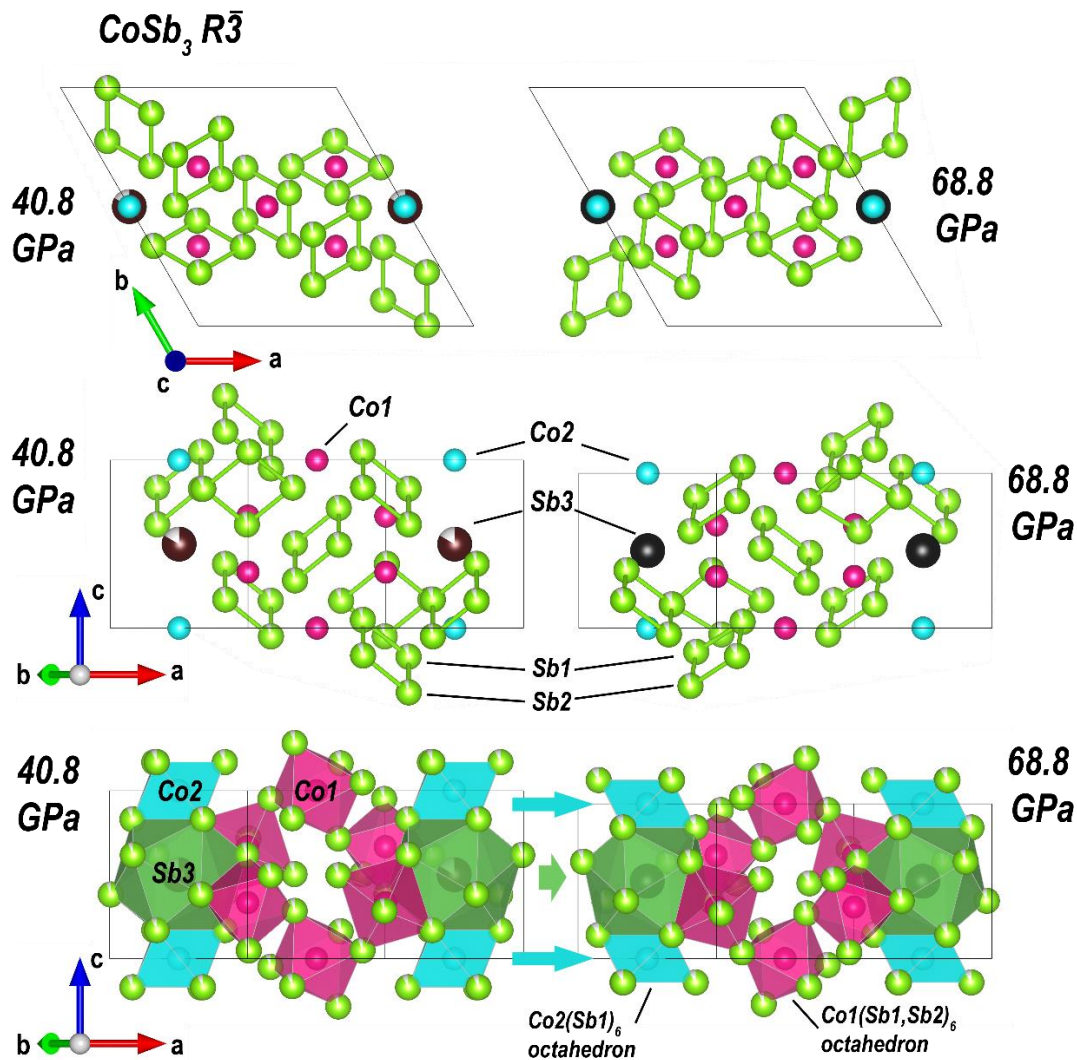

**Figure 12** Illustration of $R\bar{3}$ phase compression driven changes. We compare data for structures solved at 40.8(2) and at 68.8(2) GPa. Individual crystallographic sites are indicated together with the unit cell vectors for each of the projections. By black color we indicate the Sb atoms populating the dodecahedral sites. Note the full occupancy of these sites at the 68.8(2) GPa. The structural solution at the highest pressure has the R-factor *R1=10.5%*. Using solid green cylinders, we indicate the shortest Sb-Sb bond distances (see also **Figure 9**). The most striking feature is the distortion of $Co2(Sb1)_6$ light blue octahedra squashed between dodecahedral sites along *c*-axis of $R\bar{3}$. The dark green dodecahedral sites together with the light blue octahedra form vertical channels. In comparison to the light blue ones, the pink octahedra remain less distorted.

**References:**

[1] H. Rakoto, E. Arushanov, M. Respaud, J. M. Broto, J. Leotin, C. Kloc, E. Bucher, and S. Askenazy, *Shubnikov–de Haas Oscillations in CoSb$_3$ Single Crystals under High Magnetic Fields*, Physica B Condens Matter **246–247**, 528 (1998).

[2] D. T. Morelli, T. Caillat, J. P. Fleurial, A. Borshchevsky, J. Vandersande, B. Chen, and C. Uher, *Low-Temperature Transport Properties of p-Type CoSb$_3$*, Phys Rev B **51**, 9622 (1995).

[3] J. E. F. S. Rodrigues, J. Gainza, F. Serrano-Sánchez, C. Marini, Y. Huttel, N. M. Nemes, J. L. Martínez, and J. A. Alonso, *Atomic Structure and Lattice Dynamics of CoSb$_3$ Skutterudite-Based Thermoelectrics*, Chemistry of Materials **34**, 1213 (2022).

[4] T. Dahal, Q. Jie, G. Joshi, S. Chen, C. Guo, Y. Lan, and Z. Ren, *Thermoelectric Property Enhancement in Yb-Doped n-Type Skutterudites Yb$_x$Co$_4$Sb$_{12}$*, Acta Mater **75**, 316 (2014).

[5] Z. Y. Liu, J. L. Zhu, X. Tong, S. Niu, and W. Y. Zhao, *A Review of CoSb$_3$-Based Skutterudite Thermoelectric Materials*, Journal of Advanced Ceramics **9**, 647 (2020).

[6] T. Caillat, A. Borshchevsky, and J. P. Fleurial, *Properties of Single Crystalline Semiconducting CoSb$_3$*, J. Appl. Phys. **80**, 4442 (1996).

[7] Y. Tang, Z. M. Gibbs, L. A. Agapito, G. Li, H. S. Kim, M. B. Nardelli, S. Curtarolo, and G. J. Snyder, *Convergence of Multi-Valley Bands as the Electronic Origin of High Thermoelectric Performance in CoSb$_3$ Skutterudites*, Nature Materials 2015 14:12 **14**, 1223 (2015).

[8] S. Li, X. Jia, and H. Ma, *First-Principles Investigation of Electronic Structure and Transport Properties of CoSb$_3$ under Different Pressures*, Chem Phys Lett **549**, 22 (2012).

[9] A. C. Kraemer, M. R. Gallas, J. A. H. Da Jornada, and C. A. Perottoni, *Pressure-Induced Self-Insertion Reaction in CoSb$_3$*, Phys Rev B Condens Matter Mater Phys **75**, 024105 (2007).

[10] X. Li, Q. Zhang, Y. Kang, C. Chen, L. Zhang, D. Yu, Y. Tian, and B. Xu, *High Pressure Synthesized Ca-Filled CoSb$_3$ Skutterudites with Enhanced Thermoelectric Properties*, J Alloys Compd **677**, 61 (2016).

[11] F. Miotto, C. A. Figueirêdo, G. R. Ramos, C. L. G. Amorim, M. R. Gallas, and C. A. Perottoni, *Antimony Desinsertion Reaction from Sb$_x$CoSb$_{3-x}$*, J Appl Phys **110**, (2011).

[12] R. Viennois, T. Kume, M. Komura, L. Girard, A. Haidoux, J. Rouquette, and M. M. Koza, *Raman-Scattering Experiments on Unfilled Skutterudite CoSb$_3$ under High Pressure and High Temperature*, Journal of Physical Chemistry C **124**, 23004 (2020).

[13] K. Matsui, J. Hayashi, K. Akahira, K. Ito, K. Takeda, and C. Sekine, *Pressure-Induced Irreversible Isosymmetric Transition of TSb$_3$ (T=Co, Rh and Ir)*, J. Phys. Conf. Ser **215**, 012005 (2010).

[14] K. Glazyrin, N. Miyajima, J. S. Smith, and K. K. M. Lee, *Compression of a Multiphase Mantle Assemblage: Effects of Undesirable Stress and Stress Annealing on the Iron Spin State Crossover in Ferropericlase*, J Geophys Res Solid Earth **121**, 3377 (2016).

[15] W. Xu, W. Dong, S. Layek, M. Shulman, K. Glazyrin, E. Bykova, M. Bykov, M. Hanfland, M. P. Pasternak, I. Leonov, E. Greenberg, and G. Kh. Rozenberg, *Pressure-Induced High-Spin/Low-Spin Disproportionated State in the Mott Insulator $FeBO_3$*, Scientific Reports 2022 12:1 **12**, 1 (2022).

[16] see Supplementary information at [URL] showing additional experimental detail, *Supplementary Information*. CIF files are also submitted as a part of Supplementary.

[17] J. Miao, T. Ishikawa, I. K. Robinson, and M. M. Murnane, *Beyond Crystallography: Diffractive Imaging Using Coherent X-Ray Light Sources*, Science (1979) **348**, 530 (2015).

[18] I. K. Robinson, I. A. Vartanyants, G. J. Williams, M. A. Pfeifer, and J. A. Pitney, *Reconstruction of the Shapes of Gold Nanocrystals Using Coherent X-Ray Diffraction*, Phys Rev Lett **87**, 195505 (2001).

[19] M. A. Pfeifer, G. J. Williams, I. A. Vartanyants, R. Harder, and I. K. Robinson, *Three-Dimensional Mapping of a Deformation Field inside a Nanocrystal*, Nature 2006 442:7098 **442**, 63 (2006).

[20] W. Yang, X. Huang, R. Harder, J. N. Clark, I. K. Robinson, and H. K. Mao, *Coherent Diffraction Imaging of Nanoscale Strain Evolution in a Single Crystal under High Pressure*, Nature Communications 2013 4:1 **4**, 1 (2013).

[21] I. Robinson and R. Harder, *Coherent X-Ray Diffraction Imaging of Strain at the Nanoscale*, Nature Materials 2009 8:4 **8**, 291 (2009).

[22] K. Glazyrin, S. Khandarkhaeva, T. Fedotenko, W. Dong, D. Laniel, F. Seiboth, A. Schropp, J. Garrevoet, D. Brückner, G. Falkenberg, A. Kubec, C. David, M. Wendt, S. Wenz, L. Dubrovinsky, N. Dubrovinskaia, H. P. Liermann, *Sub-Micrometer Focusing Setup for High-Pressure Crystallography at the Extreme Conditions Beamline at PETRA III*, J. Synchrotron. Radiat. **29**, 654 (2022).

[23] M. Pillaca, O. Harder, W. Miller, and P. Gille, *Forced Convection by Inclined Rotary Bridgman Method for Growth of $CoSb_3$ and $FeSb_2$ Single Crystals from Sb-Rich Solutions*, J Cryst Growth **475**, 346 (2017).

[24] R. Boehler and K. De Hantsetters, *New Anvil Designs in Diamond-Cells*, International Journal of High Pressure Research **24**, 391 (2004).

[25] I. Kantor, V. Prakapenka, A. Kantor, P. Dera, A. Kurnosov, S. Sinogeikin, N. Dubrovinskaia, and L. Dubrovinsky, *BX90: A New Diamond Anvil Cell Design for X-Ray Diffraction and Optical Measurements*, Review of Scientific Instruments **83**, 125102 (2012).

[26] A. Dewaele, M. Torrent, P. Loubeyre, and M. Mezouar, *Compression Curves of Transition Metals in the Mbar Range: Experiments and Projector Augmented-Wave Calculations*, Phys Rev B Condens Matter Mater Phys **78**, 104102 (2008).

[27] Y. Fei, A. Ricolleau, M. Frank, K. Mibe, G. Shen, and V. Prakapenka, *Toward an Internally Consistent Pressure Scale*, Proceedings of the National Academy of Sciences **104**, 9182 (2007).

[28] A. Dewaele, A. D. Rosa, N. Guignot, D. Andrault, J. E. F. S. Rodrigues, and G. Garbarino, *Stability and Equation of State of Face-Centered Cubic and Hexagonal Close Packed Phases of Argon under Pressure*, Scientific Reports 2021 11:1 **11**, 1 (2021).

[29] H.-P. Liermann, Z. Konôpková, W. Morgenroth, K. Glazyrin, J. Bednarčik, E. E. McBride, S. Petitgirard, J. T. Delitz, M. Wendt, Y. Bican, A. Ehnes, I. Schwark, A. Rothkirch, M. Tischer, J. Heuer, H. Schulte-Schrepping, T. Kracht, and H. Franz, *The Extreme Conditions Beamline P02.2 and the Extreme Conditions Science Infrastructure at PETRA III*, J Synchrotron Radiat **22**, 908 (2015).

[30] C. Prescher and V. B. Prakapenka, *DIOPTAS: A Program for Reduction of Two-Dimensional X-Ray Diffraction Data and Data Exploration*, Http://Dx.Doi.Org/10.1080/08957959.2015.1059835 **35**, 223 (2015).

[31] Rigaku Crysalis Pro v. 171.40, *CrysAlis PRO*.

[32] V. Petricek, M. Dusek, and L. Palatinus, *JANA2006. The Crystallographic Computing System*.

[33] O. V. Dolomanov, L. J. Bourhis, R. J. Gildea, J. A. K. Howard, and H. Puschmann, *OLEX2: A Complete Structure Solution, Refinement and Analysis Program*, J. Appl. Cryst. **42**, 339 (2009).

[34] G. M. Sheldrick, *A Short History of SHELX*, Acta Cryst. A**, 64**, 112 (2008).

[35] P. Müller, *Crystal Structure Refinement: A Crystallographers Guide to SHELXL* (Oxford University Press, 2006).

[36] K. Momma and F. Izumi, *VESTA 3 for Three-Dimensional Visualization of Crystal, Volumetric and Morphology Data*, Journal of Apllied Crystallography **44**, 1272 (2011).

[37] D. C. Palmer, *CrystalMaker X*, 10.8.1.

[38] H. Xu, Z. Ren, and M. Sprung, *PyCXIM: Python Scripts for Coherent X-Ray Imaging Methods*, (unpublished).

[39] S. Marchesini, H. He, N. Chapman, P. Hau-Riege, A. Noy, R. Howells, U. Weierstall, and H. Spence, *X-Ray Image Reconstruction from a Diffraction Pattern Alone*, Phys Rev B **68**, 140101R (2003).

[40] T. Rosenqvist, *Magnetic and Crystallographic Studies on the Higher Antimonies of Iron, Cobalt and Nickel*, Acta Metallurgica **1**, 761 (1953).

[41] S. Klotz, J. C. Chervin, P. Munsch, and G. Le Marchand, *Hydrostatic Limits of 11 Pressure Transmitting Media*, J Phys D Appl Phys **42**, 075413 (2009).

[42] J. Q. Li, X. W. Feng, W. A. Sun, W. Q. Ao, F. S. Liu, and Y. Du, *Solvothermal Synthesis of Nano-Sized Skutterudite $Co_{4-x}Fe_xSb_{12}$ Powders*, Mater Chem Phys **112**, 57 (2008).

[43] C. S. Barrett, P. Cucka, and K. Haefner, *The Crystal Structure of Antimony at 4.2, 78 and 298° K*, Acta Cryst., **16**, 451 (1963).

[44] D. Schiferl and D. Schiferlt, *50-kilobar Gasketed Diamond Anvil Cell for Single-crystal X-ray Diffractometer Use with the Crystal Structure of Sb up to 26 Kilobars as a Test Problem*, Review of Scientific Instruments **48**, 24 (1977).

[45] D. Schiferl, D. T. Cromer, and J. C. Jamieson, *Structure Determinations on Sb up to $85 \times 10^2$ MPa*, Acta Cryst. B, **37**, 807 (1981).

[46] U. Schwarz, L. Akselrud, H. Rosner, A. Ormeci, Y. Grin, and M. Hanfland, *Structure and Stability of the Modulated Phase Sb-II*, Phys Rev B **67**, 214101 (2003).

[47] O. Degtyareva, M. I. Mcmahon, and R. J. Nelmes, *High-Pressure Structural Studies of Group-15 Elements*, High Press Res **24**, 319 (2004).

[48] S. Lee, H. Xu, H. Xu, and J. Neuefeind, *Crystal Structure of Moganite and Its Anisotropic Atomic Displacement Parameters Determined by Synchrotron X-Ray Diffraction and X-Ray/Neutron Pair Distribution Function Analyses*, Minerals 2021, Vol. 11, Page 272 **11**, 272 (2021).

[49] L. Hammerschmidt, S. Schlecht, and B. Paulus, *Electronic Structure and the Ground-State Properties of Cobalt Antimonide Skutterudites: Revisited with Different Theoretical Methods*, Physica Status Solidi (a) **210**, 131 (2013).

[50] L. Wu, Y. Sun, G. Z. Zhang, and C. X. Gao, *Pressure-Induced Improvement of Seebeck Coefficient and Thermoelectric Efficiency of $CoSb_3$*, Mater Lett **129**, 68 (2014).

[51] V. V. Brazhkin, D. R. Dmitriev, and R. N. Voloshin, *Metallization of $Mg_3Bi_2$ under Pressure in Crystalline and Liquid State*, Phys Lett A **193**, 102 (1994).

[52] C. W. Nicholson, F. Petocchi, B. Salzmann, C. Witteveen, M. Rumo, G. Kremer, O. Ivashko, F. O. Von Rohr, P. Werner, and C. Monney, *Gap Collapse and Flat Band Induced by Uniaxial Strain in $1T-TaS_2$*, Phys Rev B **109**, 035167 (2024).

# SUPPLEMENTARY

## Resolving the pressure induced 'self-insertion' in skutterudite $CoSb_3$

Bihan Wang[1,2], Anna Pakhomova[1,3], Saiana Khandarkhaeva[4], Mirtha Pillaca[5], Peter Gille[5], Zhe Ren[1], Dmitry Lapkin[1,6], Dameli Assalauova[1], Pavel Alexeev[1], Ilya Sergeev[1], Satishkumar Kulkarni[1], Tsu-Chien Weng[2], Michael Sprung[1], Hanns-Peter Liermann[1], Ivan A. Vartanyants[1], Konstantin Glazyrin[1,*]

[1] Photon Science, Deutsches Elektronen-Synchrotron DESY, Notkestr. 85, 22607 Hamburg, Germany; [2] Center for Transformative Science, Shanghai Technical University, Shanghai, China; [3] European Synchrotron Radiation Facility, avenue des Martyrs 71, 38043 Grenoble, France; [4] Bayerisches Geoinstitut, University of Bayreuth, Universitätstr. 30, 95440 Bayreuth, Germany; [5] Ludwig-Maximilians-Universität München, Department of Earth and Environmental Sciences, Crystallography Section, Luisenstr. 37, 80333 München, Germany; [6] Current address: Institute of Applied Physics, University of Tübingen, Auf der Morgenstelle 10, 72076 Tübingen, Germany

## 1. Crystal structure of $CoSb_3$ at ambient conditions.

At ambient conditions and after quenching from exceeding the pressures of the "self-insertion" crossover, the crystal structure of $CoSb_3$ is cubic $Im\bar{3}$ (S.G. 204). It is shown in **Figure S1** prepared in VESTA [1].

The structure is formed by corner sharing $CoSb_6$ octahedrons. Considering the shortest distances, the Co-Sb bonds within the octahedrons and the Sb-Sb bonds (highlighted in green and blue, respectively) can be identified as the characteristic bond distances within the structure. Although, the octahedrons seem to be perfect, and indeed, all Co-Sb bonds are of the same value, the octahedrons are slightly distorted even at ambient as can be seen after an inspection of the octahedra angles.

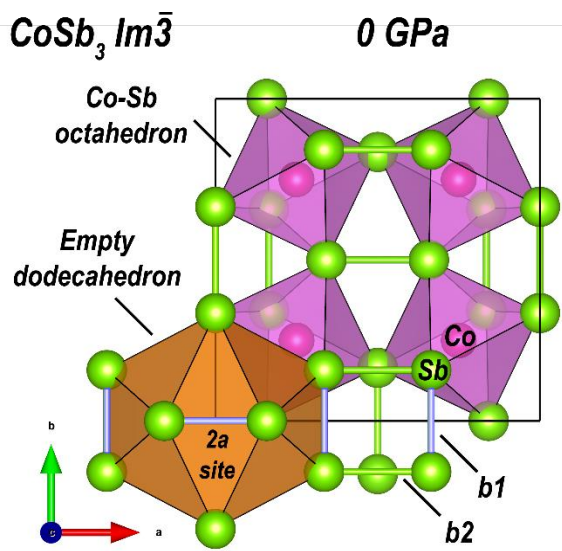

**Figure S1** Structure of $CoSb_3$ at ambient conditions. The material at ambient conditions is built from a framework of $CoSb_6$ corner sharing octahedrons, additionally bound by Sb-Sb bonds highlighted in plum blue (b1) and green color (b2) color. The empty $2a$-Wykhoff site of $Im\bar{3}$ is indicated by orange dodecahedron incorporating into its shell the shortest Sb-Sb bonds (blue color). The dodecahedron site-center can be filled by Sb atom with occupancy disorder ranging from 0 to almost 1 as a result of compression driven 'self-insertion' process. Quenching from pressures exceeding the start of "self-insertion" to ambient, may result in residual Sb $2a$-site occupancy. See the main text and the references cited within for additional discussion.

## 2. Comparison with previous studies

Considering the previous studies, we can compare our single crystal data (selected loadings with He, Ne pressure transmitting medium – PTM) with powder data from Kraemer et al. and Matsui et al. (methanol-ethanol mixture PTM) [2,3] (**Figure S2**). The data of Kraemer et al. includes different loadings indicated by the same green symbols. Inspecting powder data, we observe significant scattering at pressures above 10 GPa resulting in ambiguity of attribution of the 'self-insertion' crossover and determination of its starting pressure. At the same time, our single crystal data collected in He and Ne PTM looks more consistent and indicate a highly low probability for PTM atom to be incorporated into structural dodecahedral voids.

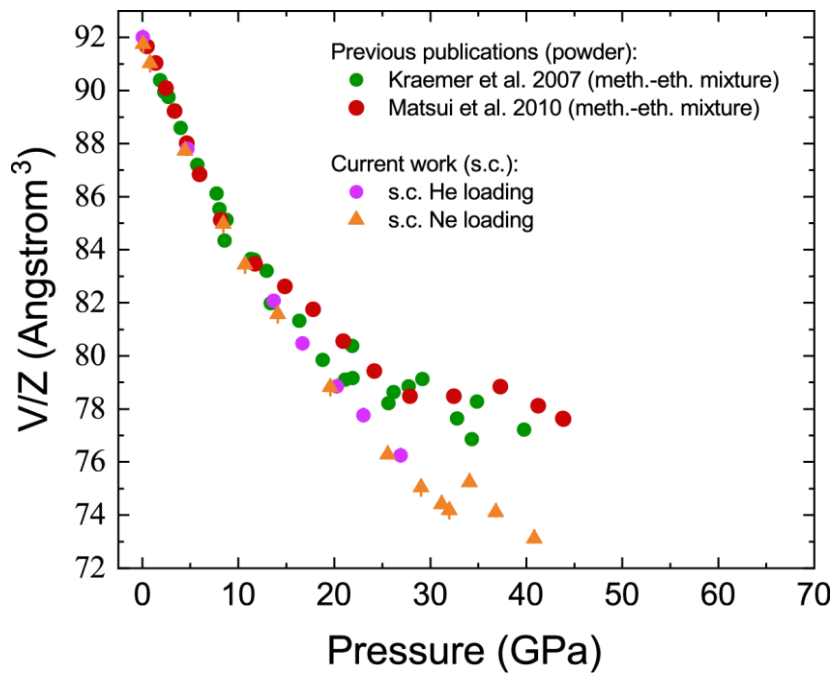

**Figure S2** Comparison of literature data (volume of a unit cell V normalized to a number of formula units per unit cell Z) collected on powder material [2,3] and loaded with methanol-ethanol mixture with our single crystal data collected in He and Ne PTM. The data of Kraemer et al. includes different loadings merged under the same symbol for the sake of comparison. We note here that methanol-ethanol mixture is considered to be quasihydrostatic up to 10-11 GPa [4]. Our data collected in much more hydrostatic He and Ne reproduce the literature data below 10 GPa and show consistency above 10 GPa.

## 3. Sample preparation and representative microphotographs

For each conventional single crystal (CSC) loading we tried to select a small crystal to minimize possible effects of undesirable stresses [5]. In **Figure S3** we show microphotographs of some of two loadings prior to loading of sample chambers with pressure transmitting media.

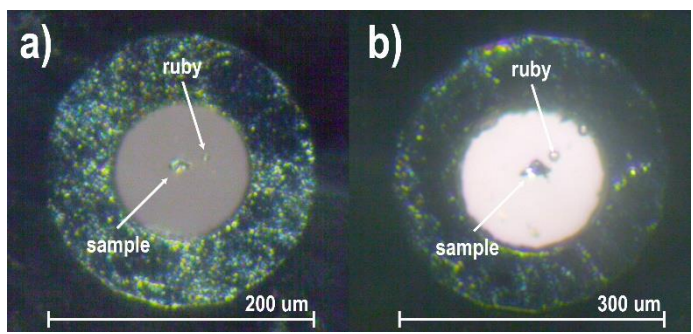

**Figure S3** Micro photographs taken prior to sample chamber gas loading with Ne (a) and Ar (b). $CoSb_3$ sample pieces and rubies are indicated by the arrows.

Preparation of sub-micron samples for single crystal X-ray diffraction with high resolution (HSC, large sample-detector-distance) was even more laborious. The snapshots indicating te important steps are shown in from electron and visible light microscopes in Figure S4.

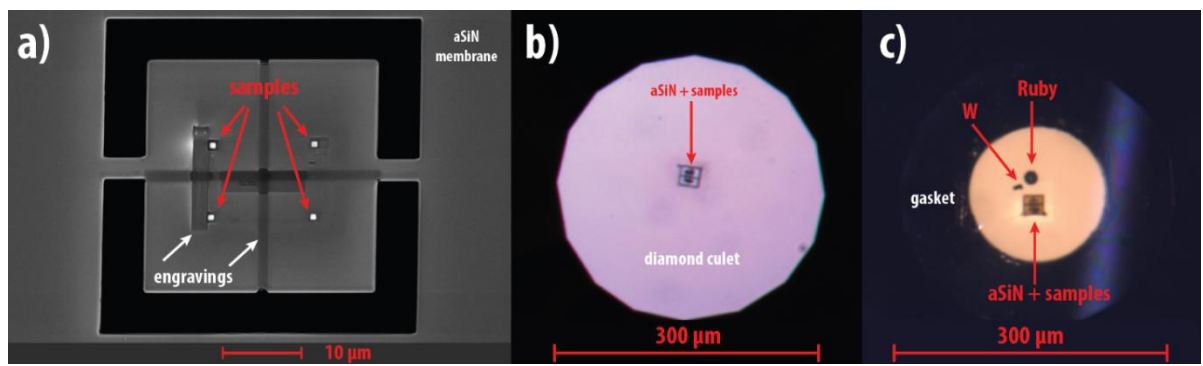

**Figure S4** Microphotographs from a) a dual beam focused ion beam (FIB) milling machine [6] and visible light microscope b) and c). We indicate the 2 µm thick amorphous silicon nitride (aSiN) membrane, FIB milled samples and additional elements of sample chamber, including Re gasket, diamond culet, tungsten piece and a ruby pressure marker. Additional markings engraved to the aSiN membrane enabled faster identification of $CoSb_3$ sample positions.

Quality of the HSC samples and their initial orientation with respect to the hosting diamond anvil cell was confirmed by means of conventional X-ray diffraction conducted using 0.2900 Å X-ray beam focused to 3×8 µm² beam at P02.2 station of PETRA-III, DESY [7]. The corresponding 2D diffraction patterns obtained using Perkin Elmer XRD 1621 are shown in Figure S5. By the standards of conventional single crystal diffraction, the collected patterns indicate high crystal quality as produced by FIB. Data collection at P02.2 was essential for pre-orientation of the same diamond anvil cell prior to its mounting at P10 of PETRA-III where HSC signal was collected.

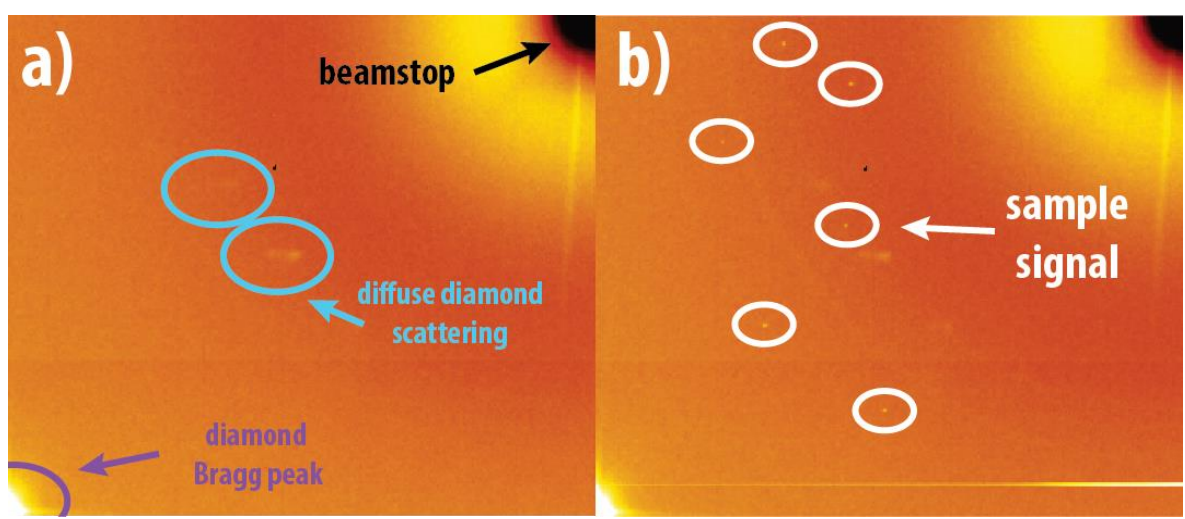

**Figure S5** Snapshots from 2D diffraction patterns collected a) outside the sub-micron crystal position and b) at the crystal. These patterns correspond to the patterns collected during a continuous rotation of sub-micron samples mounted in a diamond anvil cell within 40° of DAC aperture (e.g. ±20° of rotation with respect to a vertical ω axis). In the left panel we show indicative features contributing to the sample chamber environment, while in the right panel we indicate diffraction spots coming from a sample crystal. Tight diffraction spots

filling-in 1-2 pixels of Perkin-Elmer detector (200 µm pixel size) indicate exceptional crystal quality by the standards of conventional high-pressure beamlines. The horizontal line appearing in b) is attributed to an artifact of detector related to the detector pixels oversaturation by diamond anvil Bragg reflection. The small barely visible ring in b) is attributed to platinum sputtering fixing the aSiN membrane to the holding diamond anvil.

## 4. Observations with conventional single crystal X-ray diffraction under high pressure conditions

In **Figure S6** we show selected X-ray diffraction patterns illustrating compression path of a single crystal of $CoSb_3$ in Ne pressure medium. As can be seen in **Figure S5**a, we started with a perfect crystal, however at pressure of 31-33 GPa we see a significant broadening of the crystal reflections indicating ongoing process of 'self-insertion'. With subsequent compression the width of the peaks increases and the ratio of signal-to-noise decreases, respectively. This affects the results of the data analysis. Similar are the observations if compressing in Ar.

Our analysis indicates that above the region of 'self-insertion', the structure changes to $R\bar{3}$. This conclusion is supported by indexing of single crystal data conducted with CRYSALIS PRO [8] as well as by line broadening and eventual splitting of the Bragg reflections visible on 1D diffraction profiles (e.g. **Figure S6**). Analysis of the $Im\bar{3}$ (013) reflection and its variation under compression indicates important parts of the process with $Im\bar{3}$ (013)→$R\bar{3}$ (341) & (322) as a result of group-subgroup transformation (**Figure S7**). Analysis was complicated by an evident gradual *c/a* lattice parameter ratio change with $R\bar{3}$ (322) sitting very close to (341) and only slightly affecting the width of lines attributed to $Im\bar{3}$ (013) at lower pressures at early stages of the $Im\bar{3} \rightarrow R\bar{3}$ transition.

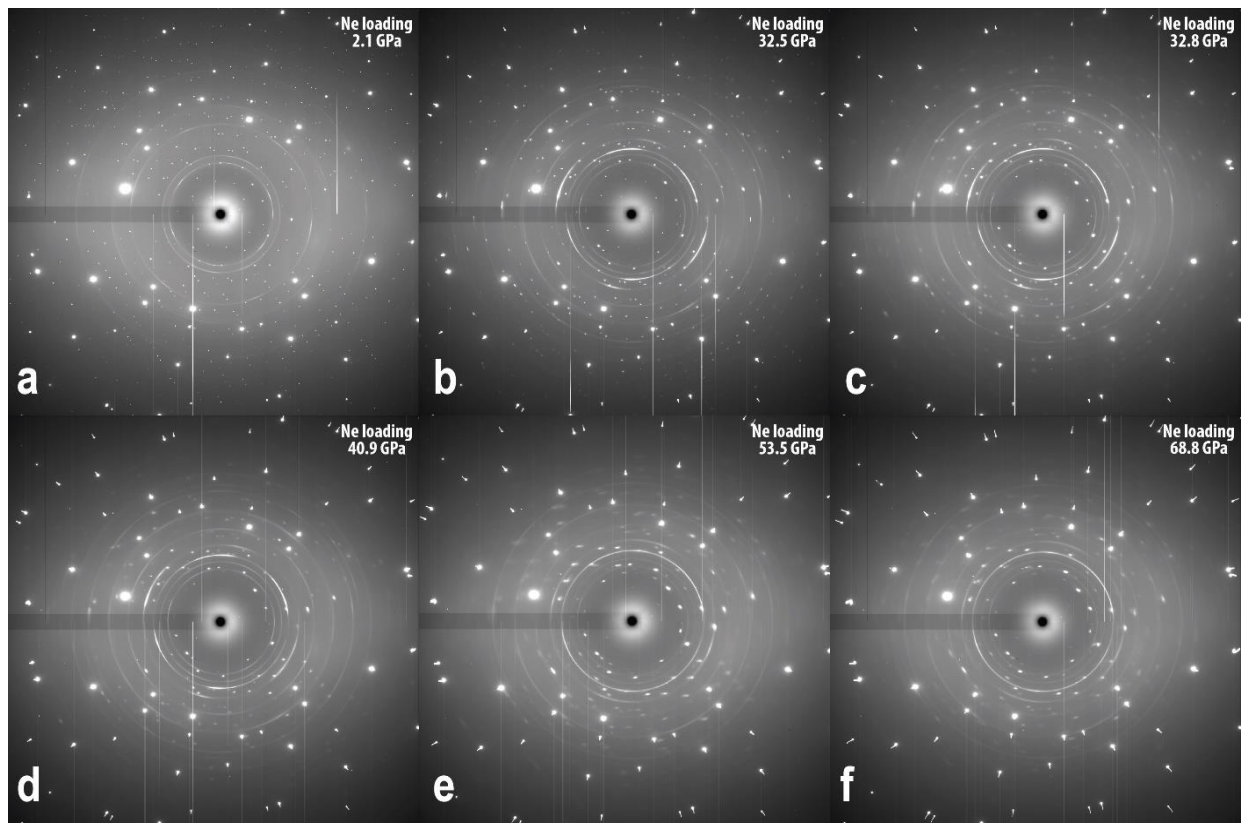

**Figure S6** X-ray diffraction patterns collected in Ne pressure medium under compression. We indicate the corresponding pressures. Images were collected during a continuous rotation of the sample centered under X-ray beam between from -20° to 20°. In addition to sample signal we observe signal from Re gasket, Ne pressure medium and Bragg spots of diamond (strongest reflections). Vertical lines seen at the images are attributed to artifacts of Perkin Elmer XRD 1621 detector readout related to pixel area oversaturation by diamond Bragg reflections and read-out synchronization.

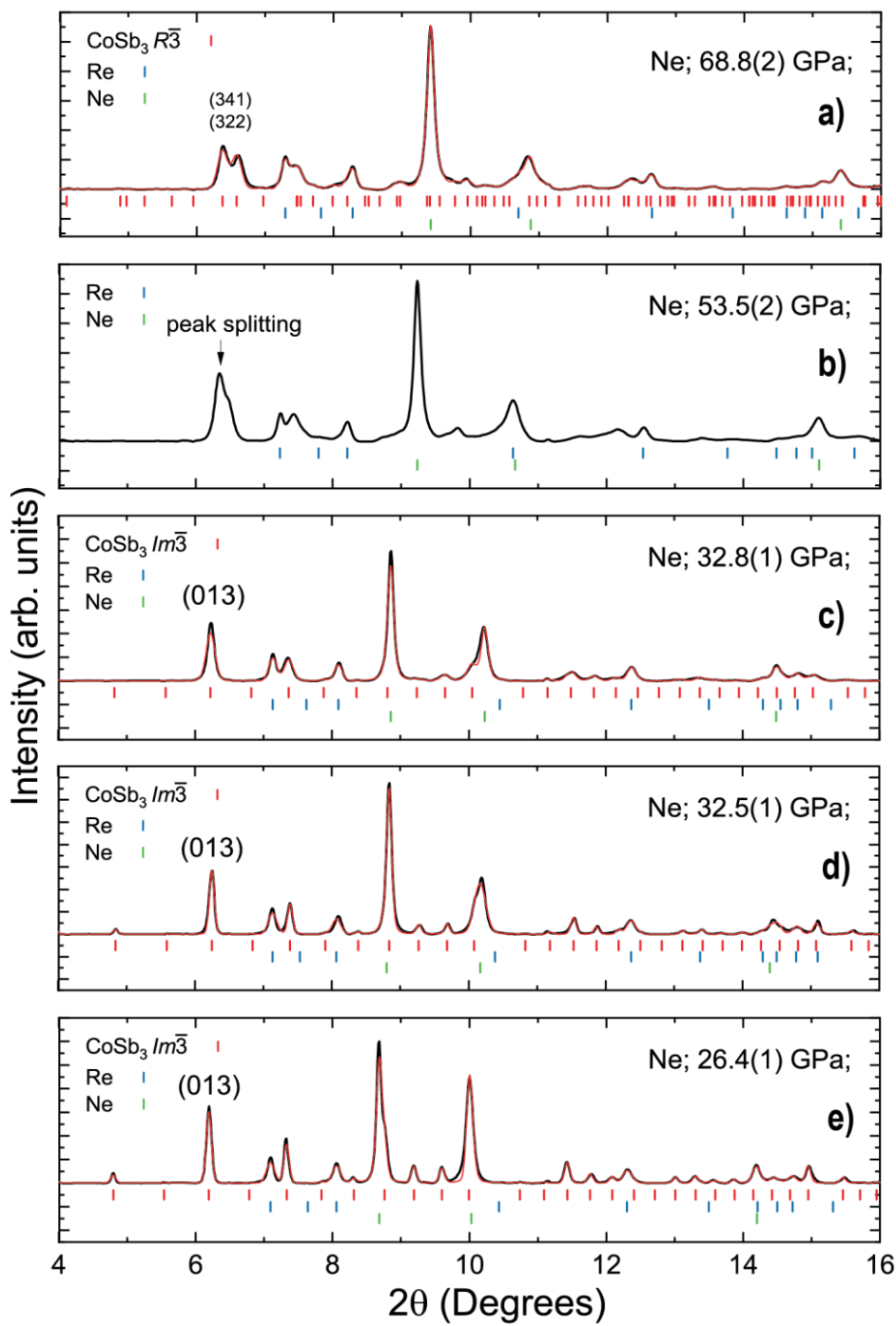

**Figure S7** Selected diffractograms CoSb$_3$ compression in Ne pressure medium sorted from highest pressure to the lowest. The patterns correspond to the 2D patterns shown in **Figure S5** with filtering of diamond reflections, artifacts of detector readout, etc. Most of the panels show Le-Bail profile fitting supporting our hypotheses. We note the significant broadening of lines coinciding with a small pressure step between c) and d) indicating the ongoing process of 'self-insertion'. Compression above 40 GPa reveals clear evidence of Bragg peak splitting which we attribute to transformation of CoSb$_3$ $Im\bar{3}$ → $R\bar{3}$ as indicated in a). The hypothesis of $R\bar{3}$ is additionally supported by indexing of single crystal data collected at 68.8(2) GPa. Considering the panel b), we deliberately hide the Le-Bail fitting of the pattern with a purpose to show substantial difficulties for indexing if using powder diffraction signal alone (e.g. peak broadening, peak overlapping, etc).

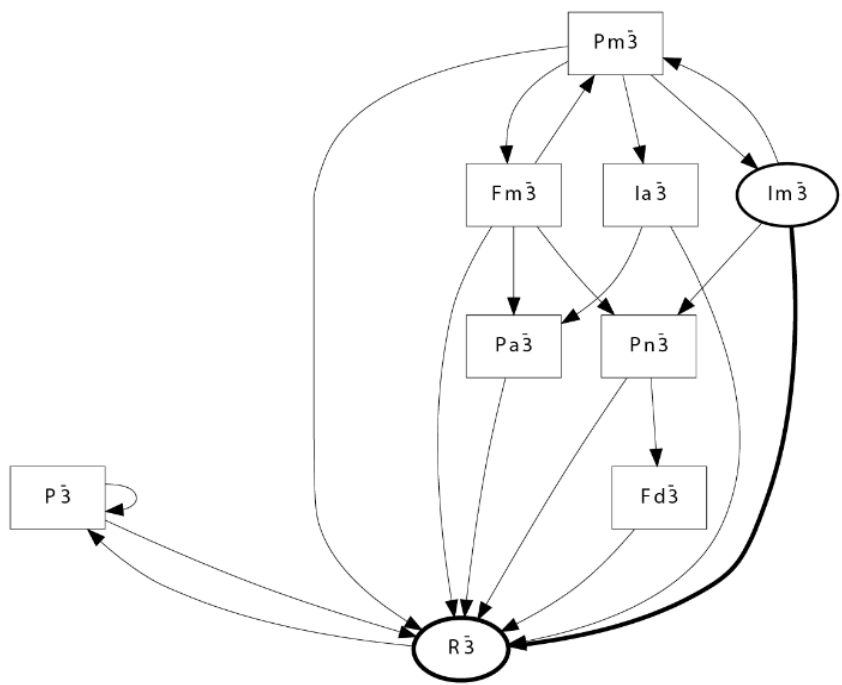

**Figure S8** Graph illustrating a group-subgroup relation between $Im\bar{3}$ and $R\bar{3}$ space groups. Plotted after Refs. [9,10]

## 5. X-ray diffraction signal at high resolution

In **Figure S9** we show evolution of other sub-micron particles and their $Im\bar{3}$ (013) Bragg peak as collected at the P10 beamline of PETRA III, DESY. These particles were hosted by the same substrate as indicated in **Figure S4**. It is clear that the signal of the Bragg spot varies from a particle to a particle. The contrast between their signal could be partially attributed to the process of particle preparation from a larger single crystalline lamella using the "top-to-bottom" approach and employing the FIB machine of DESY NanoLab for this task. We note that 3D shapes shown below cannot be used to judge the FWHM values corresponding to the Bragg spot.

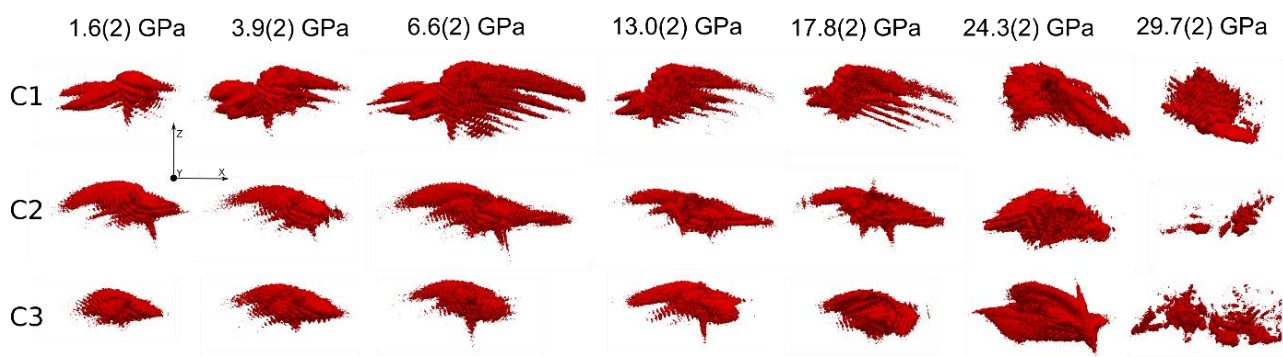

**Figure S9** Illustration of sub-micron particles signal obtained by measuring $Im\bar{3}$ (013) Bragg peak.

## 6. Elements of crystal chemistry

In **Figure S9** we show evolution of structural polyhedral volumes as a function of pressure.

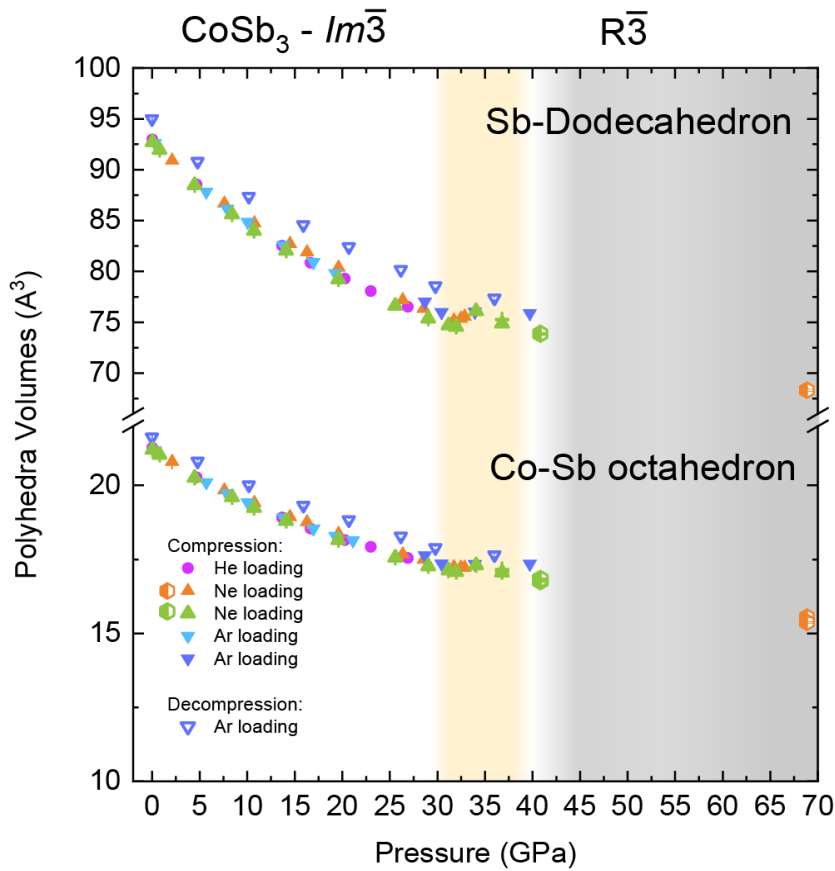

**Figure S10** Variation of polyhedral volume contributing to the framework of $CoSb_3$. Solid circles and triangles correspond to the cubic $Im\bar{3}$ phase. At the same time hexagons correspond to the trigonal $R\bar{3}$ phase (region of the grey shading). Yellow shading is attributed to the region of 'self-insertion'. Open inverted triangles indicate data collected on decompression in Ar PTM, i.e. during the same loading as loading indicated by solid blue triangles. If error bars are invisible, they are the size of the symbols or below.

## 7. References

[1] K. Momma and F. Izumi, *VESTA 3 for Three-Dimensional Visualization of Crystal, Volumetric and Morphology Data*, Journal of Apllied Crystallography **44**, 1272 (2011).

[2] A. C. Kraemer, M. R. Gallas, J. A. H. Da Jornada, and C. A. Perottoni, *Pressure-Induced Self-Insertion Reaction in $CoSb_3$*, Physical Review B, **75**, 024105 (2007).

[3] K. Matsui, J. Hayashi, K. Akahira, K. Ito, K. Takeda, and C. Sekine, *Pressure-Induced Irreversible Isosymmetric Transition of $TSb_3$ (T=Co, Rh and Ir)*, J Phys Conf Ser **215**, 012005 (2010).

[4] S. Klotz, J. C. Chervin, P. Munsch, and G. Le Marchand, *Hydrostatic Limits of 11 Pressure Transmitting Media*, Journal of Physics D: Applied Physics **42**, 075413 (2009).

[5] K. Glazyrin, N. Miyajima, J. S. Smith, and K. K. M. Lee, *Compression of a Multiphase Mantle Assemblage: Effects of Undesirable Stress and Stress Annealing on the Iron Spin State Crossover in Ferropericlase*, J Geophys Res Solid Earth **121**, 3377 (2016).

[6] A. Stierle, T. F. Keller, H. Noei, V. Vonk, and R. Roehlsberger, *DESY NanoLab*, Journal of Large-Scale Research Facilities JLSRF **2**, A76 (2016).

[7] H.-P. Liermann et al., *The Extreme Conditions Beamline P02.2 and the Extreme Conditions Science Infrastructure at PETRA III*, J Synchrotron Radiat **22**, 908 (2015).

[8] Rigaku Crysalis Pro v. 171.40, *CrysAlis PRO*.

[9] M. I. Aroyo, J. M. Perez-Mato, C. Capillas, E. Kroumova, S. Ivantchev, G. Madariaga, A. Kirov, and H. Wondratschek, *Bilbao Crystallographic Server: I. Databases and Crystallographic Computing Programs*, Zeitschrift Fur Kristallographie **221**, 15 (2006).
[10] M. I. Aroyo, A. Kirov, C. Capillas, J. M. Perez-Mato, and H. Wondratschek, *Bilbao Crystallographic Server. II. Representations of Crystallographic Point Groups and Space Groups*, Acta Cryst. A, **62**, 115 (2006).

## 8. Tables with selected single crystal data

### Dataset: 001. CIF File: DAC01_Ne_Im3_P01.cif.

| Compound: | $CoSb_3$ | PTM: Ne |
|---|---|---|
| Pressure: | 2.1 GPa | Temperature: 293 K |
| Wavelength: | 0.2907 Å | |
| Space group: | *Im -3*, S.G. #204 | |
| Z: | 8 | |

| Lattice parameters | | | |
|---|---|---|---|
| a, Å | b, Å | c, Å | V, Å$^3$ |
| 8.9650(2) | 8.9650(2) | 8.9650(2) | |
| α, ° | β, ° | γ, ° | |
| 90 | 90 | 90 | |

| Refinement information | | | |
|---|---|---|---|
| $R1, I>2\sigma(I)$ | 2.59 % | $\rho_{min}, \rho_{max}, e^-/Å^3$ | -1.49, 2.87 |
| $wR2, I>2\sigma(I)$ | 6.21 % | | $-17 \leq H \leq 18$ |
| $N_{par}/N_{obs}$ | 10/374 | $2.27° \leq \theta \leq 17.43°$ | $-15 \leq K \leq 16$ |
| | | | $-11 \leq L \leq 9$ |

| Structural information | | | | |
|---|---|---|---|---|
| site | *x* | *y* | *z* | Occ. |
| Co1 | 0.25 | 0.25 | 0.25 | 1 |
| Sb1 | 0 | 0.1580(1) | 0.3349(1) | 1 |

*I* – signal intensity; σ(*I*) – standard deviation of *I*; *R1, wR2* – crystallographic R-factors; $N_{par}/N_{obs}$ – ratio between number of parameters to number of observed reflections for $I>2\sigma(I)$; $\rho_{min}, \rho_{max}$ – minimum and maximum values of residual electronic density; PTM – pressure transmitting medium;

**Dataset: 002.** CIF File: DAC01_Ne_Im3_P03.cif.

| Compound: | CoSb$_3$ | PTM: Ne |
|---|---|---|
| Pressure: | 7.6 GPa | Temperature: 293 K |
| Wavelength: | 0.2907 Å | |
| Space group: | *I m -3*, S.G. #204 | |
| Z: | 8 | |

| Lattice parameters | | | |
|---|---|---|---|
| a, Å | b, Å | c, Å | V, Å$^3$ |
| 8.82850(10) | 8.82850(10) | 8.82850(10) | |
| α, ° | β, ° | γ, ° | |
| 90 | 90 | 90 | |

| Refinement information | | | |
|---|---|---|---|
| *R1, I>2σ(I)* | 2.44 % | $\rho_{min}, \rho_{max},$ e$^-$/Å$^3$ | -1.25, 1.56 |
| *wR2, I>2σ(I)* | 5.69 % | | -9 ≤ H ≤ 11 |
| *N$_{par}$/N$_{obs}$* | 10/345 | 2.31°≤ θ ≤16.92° | -17 ≤ K ≤ 17 |
| | | | -15 ≤ L ≤ 16 |

| Structural information | | | | |
|---|---|---|---|---|
| site | *x* | *y* | *z* | Occ. |
| Co1 | 0.25 | 0.25 | 0.25 | 1 |
| Sb1 | 0 | 0.1581(1) | 0.3345(1) | 1 |

*I* – signal intensity; *σ(I)* – standard deviation of *I*; *R1, wR2* – crystallographic R-factors; *N$_{par}$/N$_{obs}$* – ratio between number of parameters to number of observed reflections for *I>2σ(I)*; $\rho_{min}, \rho_{max}$ – minimum and maximum values of residual electronic density; PTM – pressure transmitting medium;

**Dataset: 003.** CIF File: DAC01_Ne_Im3_P04.cif.

| Compound: | CoSb$_3$ | PTM: Ne |
|---|---|---|
| Pressure: | 10.8 GPa | Temperature: 293 K |
| Wavelength: | 0.2907 Å | |
| Space group: | *I m -3*, S.G. #204 | |
| Z: | 8 | |

| Lattice parameters | | | |
|---|---|---|---|
| a, Å | b, Å | c, Å | V, Å$^3$ |
| 8.7652(4) | 8.7652(4) | 8.7652(4) | |
| α, ° | β, ° | γ, ° | |
| 90 | 90 | 90 | |

| Refinement information | | | |
|---|---|---|---|
| *R1, I>2σ(I)* | 2.51 % | $\rho_{min}, \rho_{max},$ e$^-$/Å$^3$ | -1.42, 1.39 |
| *wR2, I>2σ(I)* | 5.31 % | | -14 ≤ H ≤ 16 |
| *N$_{par}$/N$_{obs}$* | 10/315 | 2.31° ≤ θ ≤ 17.06° | -16 ≤ K ≤ 17 |
| | | | -9 ≤ L ≤ 11 |

| Structural information | | | | |
|---|---|---|---|---|
| site | *x* | *y* | *z* | Occ. |
| Co1 | 0.25 | 0.25 | 0.25 | 1 |
| Sb1 | 0 | 0.3344(1) | 0.1581(1) | 1 |

*I* – signal intensity; *σ(I)* – standard deviation of *I*; *R1, wR2* – crystallographic R-factors; *N$_{par}$/N$_{obs}$* – ratio between number of parameters to number of observed reflections for *I>2σ(I)*; $\rho_{min}, \rho_{max}$ – minimum and maximum values of residual electronic density; PTM – pressure transmitting medium;

**Dataset: 004.** CIF File: DAC01_Ne_Im3_P05.cif.

| Compound: | CoSb$_3$ | PTM: Ne |
|---|---|---|
| Pressure: | 14.5 GPa | Temperature: 293 K |
| Wavelength: | 0.2907 Å | |
| Space group: | *Im -3*, S.G. #204 | |
| Z: | 8 | |

| Lattice parameters | | | |
|---|---|---|---|
| a, Å | b, Å | c, Å | V, Å$^3$ |
| 8.6953(6) | 8.6953(6) | 8.6953(6) | |
| α, ° | β, ° | γ, ° | |
| 90 | 90 | 90 | |

| Refinement information | | | |
|---|---|---|---|
| R1, I>2σ(I) | 4.53 % | ρ$_{min}$, ρ$_{max}$, e$^-$/Å$^3$ | -3.73, 3.01 |
| wR2, I>2σ(I) | 10.08 % | | -10 ≤ H ≤ 9 |
| N$_{par}$/N$_{obs}$ | 10/304 | 2.33° ≤ θ ≤ 16.22° | -17 ≤ K ≤ 16 |
| | | | -16 ≤ L ≤ 14 |

| Structural information | | | | |
|---|---|---|---|---|
| site | x | y | z | Occ. |
| Co1 | 0.25 | 0.25 | 0.25 | 1 |
| Sb1 | 0 | 0.1582(1) | 0.3343(1) | 1 |

*I* – signal intensity; σ(I) – standard deviation of *I*; *R1, wR2* – crystallographic R-factors; N$_{par}$/N$_{obs}$ – ratio between number of parameters to number of observed reflections for I>2σ(I); ρ$_{min}$, ρ$_{max}$ – minimum and maximum values of residual electronic density; PTM – pressure transmitting medium;

**Dataset: 005.** CIF File: DAC01_Ne_Im3_P06.cif.

| Compound: | CoSb$_3$ | PTM: Ne |
|---|---|---|
| Pressure: | 16.3 GPa | Temperature: 293 K |
| Wavelength: | 0.2907 Å | |
| Space group: | *Im -3*, S.G. #204 | |
| Z: | 8 | |

### Lattice parameters

| a, Å | b, Å | c, Å | V, Å$^3$ |
|---|---|---|---|
| 8.6687(2) | 8.6687(2) | 8.6687(2) | |
| α, ° | β, ° | γ, ° | |
| 90 | 90 | 90 | |

### Refinement information

| R1, I>2σ(I) | 2.96 % | ρ$_{min}$, ρ$_{max}$, e$^-$/Å$^3$ | -2.02, 2.19 |
|---|---|---|---|
| wR2, I>2σ(I) | 6.56 % | | -14 ≤ H ≤ 16 |
| N$_{par}$/N$_{obs}$ | 10/319 | 2.35° ≤ θ ≤ 16.51° | -16 ≤ K ≤ 17 |
| | | | -9 ≤ L ≤ 10 |

### Structural information

| site | x | y | z | Occ. |
|---|---|---|---|---|
| Co1 | 0.25 | 0.25 | 0.25 | 1 |
| Sb1 | 0 | 0.3342(1) | 0.1582(1) | 1 |

*I* – signal intensity; σ(I) – standard deviation of *I*; *R1, wR2* – crystallographic R-factors; N$_{par}$/N$_{obs}$ – ratio between number of parameters to number of observed reflections for *I>2σ(I)*; ρ$_{min}$, ρ$_{max}$ – minimum and maximum values of residual electronic density; PTM – pressure transmitting medium;

**Dataset: 006.** CIF File: DAC01_Ne_Im3_P07.cif.

| Compound: | CoSb$_3$ | PTM: Ne |
|---|---|---|
| Pressure: | 19.6 GPa | Temperature: 293 K |
| Wavelength: | 0.2907 Å | |
| Space group: | $Im\bar{3}$, S.G. #204 | |
| Z: | 8 | |

### Lattice parameters

| a, Å | b, Å | c, Å | V, Å$^3$ |
|---|---|---|---|
| 8.6092(5) | 8.6092(5) | 8.6092(5) | |
| α, ° | β, ° | γ, ° | |
| 90 | 90 | 90 | |

### Refinement information

| | | | |
|---|---|---|---|
| R1, I>2σ(I) | 4.93 % | $\rho_{min}, \rho_{max}, e^-/Å^3$ | -3.82, 4.15 |
| wR2, I>2σ(I) | 11.74 % | | -16 ≤ H ≤ 14 |
| N$_{par}$/N$_{obs}$ | 10/281 | 2.35°≤ θ ≤16.37° | -10 ≤ K ≤ 9 |
| | | | -17 ≤ L ≤ 16 |

### Structural information

| site | x | y | z | Occ. |
|---|---|---|---|---|
| Co1 | 0.25 | 0.25 | 0.25 | 1 |
| Sb1 | 0 | 0.1584(1) | 0.3343(1) | 1 |

*I* – signal intensity; σ(I) – standard deviation of I; *R1, wR2* – crystallographic R-factors; N$_{par}$/N$_{obs}$ – ratio between number of parameters to number of observed reflections for I>2σ(I); $\rho_{min}, \rho_{max}$ – minimum and maximum values of residual electronic density; PTM – pressure transmitting medium;

**Dataset: 007.** CIF File: DAC01_Ne_Im3_P08.cif.

| Compound: | CoSb$_3$ | PTM: Ne |
| --- | --- | --- |
| Pressure: | 26.3 GPa | Temperature: 293 K |
| Wavelength: | 0.2907 Å | |
| Space group: | *I m -3*, S.G. #204 | |
| Z: | 8 | |

| Lattice parameters | | | |
| --- | --- | --- | --- |
| a, Å | b, Å | c, Å | V, Å$^3$ |
| 8.5033(6) | 8.5033(6) | 8.5033(6) | |
| α, ° | β, ° | γ, ° | |
| 90 | 90 | 90 | |

| Refinement information | | | |
| --- | --- | --- | --- |
| *R1, I>2σ(I)* | 3.18 % | $\rho_{min}, \rho_{max},$ e$^-$/Å$^3$ | -1.85, 2.44 |
| *wR2, I>2σ(I)* | 7.28 % | | -16 ≤ H ≤ 17 |
| $N_{par}/N_{obs}$ | 10/279 | 2.41° ≤ θ ≤ 16.81° | -9 ≤ K ≤ 10 |
| | | | -14 ≤ L ≤ 15 |

| Structural information | | | | |
| --- | --- | --- | --- | --- |
| site | *x* | *y* | *z* | Occ. |
| Co1 | 0.25 | 0.25 | 0.25 | 1 |
| Sb1 | 0 | 0.3338(1) | 0.1584(1) | 1 |

*I* – signal intensity; *σ(I)* – standard deviation of *I*; *R1, wR2* – crystallographic R-factors; $N_{par}/N_{obs}$ – ratio between number of parameters to number of observed reflections for *I>2σ(I)*; $\rho_{min}, \rho_{max}$ – minimum and maximum values of residual electronic density; PTM – pressure transmitting medium;

**Dataset: 008.** CIF File: DAC01_Ne_Im3_P09.cif.

| Compound: | CoSb$_3$ | PTM: Ne |
|---|---|---|
| Pressure: | 28.5 GPa | Temperature: 293 K |
| Wavelength: | 0.2907 Å | |
| Space group: | $Im\text{-}3$, S.G. #204 | |
| Z: | 8 | |

| Lattice parameters | | | |
|---|---|---|---|
| a, Å | b, Å | c, Å | V, Å$^3$ |
| 8.4741(9) | 8.4741(9) | 8.4741(9) | |
| α, ° | β, ° | γ, ° | |
| 90 | 90 | 90 | |

| Refinement information | | | |
|---|---|---|---|
| $R1, I>2\sigma(I)$ | 3.00 % | $\rho_{min}, \rho_{max}, e^-/Å^3$ | -3.15, 1.93 |
| $wR2, I>2\sigma(I)$ | 6.38 % | | $-14 \leq H \leq 15$ |
| $N_{par}/N_{obs}$ | 10/265 | $2.38° \leq \theta \leq 16.62°$ | $-10 \leq K \leq 9$ |
| | | | $-15 \leq L \leq 17$ |

| Structural information | | | | |
|---|---|---|---|---|
| site | x | y | z | Occ. |
| Co1 | 0.25 | 0.25 | 0.25 | 1 |
| Sb1 | 0 | 0.1584(1) | 0.3337(1) | 1 |

*I* – signal intensity; $\sigma(I)$ – standard deviation of *I*; *R1, wR2* – crystallographic R-factors; $N_{par}/N_{obs}$ – ratio between number of parameters to number of observed reflections for $I>2\sigma(I)$; $\rho_{min}, \rho_{max}$ – minimum and maximum values of residual electronic density; PTM – pressure transmitting medium;

**Dataset: 009.** CIF File: DAC01_Ne_Im3_P10.cif.

| Compound: | CoSb$_3$ | PTM: Ne |
|---|---|---|
| Pressure: | 31.8 GPa | Temperature: 293 K |
| Wavelength: | 0.2907 Å | |
| Space group: | *I m -3*, S.G. #204 | |
| Z: | 8 | |

| Lattice parameters | | | |
|---|---|---|---|
| a, Å | b, Å | c, Å | V, Å$^3$ |
| 8.4325(8) | 8.4325(8) | 8.4325(8) | |
| α, ° | β, ° | γ, ° | |
| 90 | 90 | 90 | |

| Refinement information | | | |
|---|---|---|---|
| *R1, I>2σ(I)* | 4.01 % | $\rho_{min}, \rho_{max}, e^-/Å^3$ | -2.75, 4.42 |
| *wR2, I>2σ(I)* | 8.81 % | | -9 ≤ H ≤ 10 |
| *N$_{par}$/N$_{obs}$* | 10/235 | 2.43°≤ θ ≤16.7° | -15 ≤ K ≤ 14 |
| | | | -17 ≤ L ≤ 15 |

| Structural information | | | | |
|---|---|---|---|---|
| site | *x* | *y* | *z* | Occ. |
| Co1 | 0.25 | 0.25 | 0.25 | 1 |
| Sb1 | 0 | 0.3337(1) | 0.1583(1) | 1 |

*I* – signal intensity; *σ(I)* – standard deviation of *I*; *R1, wR2* – crystallographic R-factors; *N$_{par}$/N$_{obs}$* – ratio between number of parameters to number of observed reflections for *I>2σ(I)*; $\rho_{min}, \rho_{max}$ – minimum and maximum values of residual electronic density; PTM – pressure transmitting medium;

**Dataset: 010.** CIF File: DAC01_Ne_Im3_P11.cif.

| Compound: | CoSb$_3$ | PTM: Ne |
|---|---|---|
| Pressure: | 32.5 GPa | Temperature: 293 K |
| Wavelength: | 0.2907 Å | |
| Space group: | $Im\bar{3}$, S.G. #204 | |
| Z: | 8 | |

| Lattice parameters ||||
|---|---|---|---|
| a, Å | b, Å | c, Å | V, Å$^3$ |
| 8.4331(8) | 8.4331(8) | 8.4331(8) | |
| α, ° | β, ° | γ, ° | |
| 90 | 90 | 90 | |

| Refinement information ||||
|---|---|---|---|
| $R1, I>2\sigma(I)$ | 6.70 % | $\rho_{min}, \rho_{max}, e^-/Å^3$ | -9.34, 4.78 |
| $wR2, I>2\sigma(I)$ | 11.35 % | | $-8 \leq H \leq 10$ |
| $N_{par}/N_{obs}$ | 10/182 | $2.39° \leq \theta \leq 16.71°$ | $-15 \leq K \leq 14$ |
| | | | $-16 \leq L \leq 16$ |

| Structural information |||||
|---|---|---|---|---|
| site | x | y | z | Occ. |
| Co1 | 0.25 | 0.25 | 0.25 | 1 |
| Sb1 | 0 | 0.3338(2) | 0.1587(2) | 0.977(1) |
| Sb2 | 0 | 0 | 0 | 0.237(11) |

*I* – signal intensity; σ(I) – standard deviation of *I*; *R1, wR2* – crystallographic R-factors; $N_{par}/N_{obs}$ – ratio between number of parameters to number of observed reflections for $I>2\sigma(I)$; $\rho_{min}, \rho_{max}$ – minimum and maximum values of residual electronic density; PTM – pressure transmitting medium;

**Dataset: 011.** CIF File: DAC01_Ne_Im3_P13.cif.

| Compound: | CoSb$_3$ | PTM: Ne |
|---|---|---|
| Pressure: | 32.9 GPa | Temperature: 293 K |
| Wavelength: | 0.2907 Å | |
| Space group: | *I m -3*, S.G. #204 | |
| Z: | 8 | |

| Lattice parameters | | | |
|---|---|---|---|
| a, Å | b, Å | c, Å | V, Å$^3$ |
| 8.4331(8) | 8.4331(8) | 8.4331(8) | |
| α, ° | β, ° | γ, ° | |
| 90 | 90 | 90 | |

| Refinement information | | | |
|---|---|---|---|
| R1, I>2σ(I) | 5.47 % | ρ$_{min}$, ρ$_{max}$, e$^-$/Å$^3$ | -2.61, 2.50 |
| wR2, I>2σ(I) | 10.94 % | | -11 ≤ H ≤ 12 |
| N$_{par}$/N$_{obs}$ | 10/123 | 2.42° ≤ θ ≤ 16.12° | -7 ≤ K ≤ 8 |
| | | | -10 ≤ L ≤ 11 |

| Structural information | | | | |
|---|---|---|---|---|
| site | x | y | z | Occ. |
| Co1 | 0.25 | 0.25 | 0.25 | 1 |
| Sb1 | 0 | 0.3341(2) | 0.1588(2) | 0.939(1) |
| Sb2 | 0 | 0 | 0 | 0.694(16) |

*I* – signal intensity; σ(I) – standard deviation of *I*; *R1, wR2* – crystallographic R-factors; N$_{par}$/N$_{obs}$ – ratio between number of parameters to number of observed reflections for I>2σ(I); ρ$_{min}$, ρ$_{max}$ – minimum and maximum values of residual electronic density; PTM – pressure transmitting medium;

**Dataset: 012.** CIF File: DAC01_Ne_R3_P24.cif.

| Compound: | CoSb$_3$ | PTM: Ne |
|---|---|---|
| Pressure: | 68.8 GPa | Temperature: 293 K |
| Wavelength: | 0.2907 Å | |
| Space group: | R -3, S.G. #148 | |
| Z: | 12 | |

### Lattice parameters

| a, Å | b, Å | c, Å | V, Å$^3$ |
|---|---|---|---|
| 11.840(10) | 11.840(10) | 6.659(8) | |
| α, ° | β, ° | γ, ° | |
| 90 | 90 | 120 | |

### Refinement information

| R1, I>2σ(I) | 10.51 % | ρ$_{min}$, ρ$_{max,}$ e$^-$/Å$^3$ | -4.29, 4.33 |
|---|---|---|---|
| wR2, I>2σ(I) | 16.61 % | | -16 ≤ H ≤ 15 |
| N$_{par}$/N$_{obs}$ | 25/121 | 2.4°≤ θ ≤11.86° | -5 ≤ K ≤ 7 |
| | | | -8 ≤ L ≤ 9 |

### Structural information

| site | x | y | z | Occ. |
|---|---|---|---|---|
| Co1 | 0.5 | 0 | 0 | 1 |
| Co2 | 0.666667 | 0.333333 | -0.166667 | 1 |
| Sb1 | 0.5919(4) | 0.0406(7) | 0.2189(4) | 0.922(10) |
| Sb2 | 0.6651(4) | 0.1801(5) | 0.0087(5) | 0.900(12) |
| Sb3 | 0.333333 | 0.166667 | 0.166667 | 1.070(90) |

*I* – signal intensity; σ(I) – standard deviation of *I*; *R1, wR2* – crystallographic R-factors; N$_{par}$/N$_{obs}$ – ratio between number of parameters to number of observed reflections for I>2σ(I); ρ$_{min}$, ρ$_{max}$ – minimum and maximum values of residual electronic density; PTM – pressure transmitting medium;

**Dataset: 013.** CIF File: DAC02_Ne_Im3_P00.cif.

| Compound: | CoSb$_3$ | PTM: Ne |
|---|---|---|
| Pressure: | 0.1 GPa | Temperature: 293 K |
| Wavelength: | 0.2952 Å | |
| Space group: | $Im\text{-}3$, S.G. #204 | |
| Z: | 8 | |

| Lattice parameters | | | |
|---|---|---|---|
| a, Å | b, Å | c, Å | V, Å$^3$ |
| 9.02060(10) | 9.02060(10) | 9.02060(10) | |
| α, ° | β, ° | γ, ° | |
| 90 | 90 | 90 | |

| Refinement information | | | |
|---|---|---|---|
| R1, I>2σ(I) | 1.80 % | $\rho_{min}$, $\rho_{max}$, e$^-$/Å$^3$ | -1.34, 1.51 |
| wR2, I>2σ(I) | 4.30 % | | -16 ≤ H ≤ 16 |
| N$_{par}$/N$_{obs}$ | 10/362 | 1.87° ≤ θ ≤ 16.24° | -14 ≤ K ≤ 15 |
| | | | -11 ≤ L ≤ 9 |

| Structural information | | | | |
|---|---|---|---|---|
| site | x | y | z | Occ. |
| Co1 | 0.25 | 0.25 | 0.25 | 1 |
| Sb1 | 0 | 0.1579(1) | 0.3351(1) | 1 |

*I* – signal intensity; *σ(I)* – standard deviation of *I*; *R1, wR2* – crystallographic R-factors; *N$_{par}$/N$_{obs}$* – ratio between number of parameters to number of observed reflections for *I>2σ(I)*; *ρ$_{min}$, ρ$_{max}$* – minimum and maximum values of residual electronic density; PTM – pressure transmitting medium;

**Dataset: 014.** CIF File: DAC02_Ne_Im3_P01.cif.

| Compound: | CoSb$_3$ | PTM: Ne |
|---|---|---|
| Pressure: | 0.8 GPa | Temperature: 293 K |
| Wavelength: | 0.2952 Å | |
| Space group: | $Im\text{-}3$, S.G. #204 | |
| Z: | 8 | |

| Lattice parameters ||||
|---|---|---|---|
| a, Å | b, Å | c, Å | V, Å$^3$ |
| 8.99700(10) | 8.99700(10) | 8.99700(10) | |
| α, ° | β, ° | γ, ° | |
| 90 | 90 | 90 | |

| Refinement information ||||
|---|---|---|---|
| $R1, I>2\sigma(I)$ | 1.81 % | $\rho_{min}, \rho_{max}, e^-/Å^3$ | -1.30, 1.49 |
| $wR2, I>2\sigma(I)$ | 4.35 % | | $-16 \leq H \leq 16$ |
| $N_{par}/N_{obs}$ | 10/377 | $1.88° \leq \theta \leq 16.28°$ | $-14 \leq K \leq 15$ |
| | | | $-12 \leq L \leq 9$ |

| Structural information |||||
|---|---|---|---|---|
| site | x | y | z | Occ. |
| Co1 | 0.25 | 0.25 | 0.25 | 1 |
| Sb1 | 0 | 0.1579(1) | 0.3350(1) | 1 |

---

*I* – signal intensity; *σ(I)* – standard deviation of *I*; *R1, wR2* – crystallographic R-factors; $N_{par}/N_{obs}$ – ratio between number of parameters to number of observed reflections for $I>2\sigma(I)$; $\rho_{min}, \rho_{max}$ – minimum and maximum values of residual electronic density; PTM – pressure transmitting medium;

**Dataset: 015.** CIF File: DAC02_Ne_Im3_P02.cif.

| Compound: | CoSb$_3$ | PTM: Ne |
|---|---|---|
| Pressure: | 4.5 GPa | Temperature: 293 K |
| Wavelength: | 0.2952 Å | |
| Space group: | $Im\text{-}3$, S.G. #204 | |
| Z: | 8 | |

| Lattice parameters | | | |
|---|---|---|---|
| a, Å | b, Å | c, Å | V, Å$^3$ |
| 8.88710(10) | 8.88710(10) | 8.88710(10) | |
| α, ° | β, ° | γ, ° | |
| 90 | 90 | 90 | |

| Refinement information | | | |
|---|---|---|---|
| R1, I>2σ(I) | 2.15 % | $\rho_{min}, \rho_{max},$ e$^-$/Å$^3$ | -1.58, 2.08 |
| wR2, I>2σ(I) | 5.12 % | | -11 ≤ H ≤ 9 |
| N$_{par}$/N$_{obs}$ | 9/354 | 1.9° ≤ θ ≤ 16.5° | -14 ≤ K ≤ 13 |
| | | | -16 ≤ L ≤ 16 |

| Structural information | | | | |
|---|---|---|---|---|
| site | x | y | z | Occ. |
| Co1 | 0.25 | 0.25 | 0.25 | 1 |
| Sb1 | 0.1653(1) | 0.3420(1) | 0.5 | 1 |

---

*I* – signal intensity; σ(I) – standard deviation of *I*; *R1, wR2* – crystallographic R-factors; N$_{par}$/N$_{obs}$ – ratio between number of parameters to number of observed reflections for I>2σ(I); $\rho_{min}, \rho_{max}$ – minimum and maximum values of residual electronic density; PTM – pressure transmitting medium;

**Dataset: 016.** CIF File: DAC02_Ne_Im3_P03.cif.

| Compound: | CoSb$_3$ | PTM: Ne |
|---|---|---|
| Pressure: | 8.4 GPa | Temperature: 293 K |
| Wavelength: | 0.2952 Å | |
| Space group: | $Im\text{-}3$, S.G. #204 | |
| Z: | 8 | |

### Lattice parameters

| a, Å | b, Å | c, Å | V, Å$^3$ |
|---|---|---|---|
| 8.79280(10) | 8.79280(10) | 8.79280(10) | |
| α, ° | β, ° | γ, ° | |
| 90 | 90 | 90 | |

### Refinement information

| R1, I>2σ(I) | 2.38 % | $\rho_{min}$, $\rho_{max}$, e$^-$/Å$^3$ | -1.91, 1.82 |
|---|---|---|---|
| wR2, I>2σ(I) | 5.54 % | | -11 ≤ H ≤ 9 |
| $N_{par}/N_{obs}$ | 9/339 | 1.92°≤ θ ≤16.67° | -14 ≤ K ≤ 13 |
| | | | -16 ≤ L ≤ 16 |

### Structural information

| site | x | y | z | Occ. |
|---|---|---|---|---|
| Co1 | 0.25 | 0.25 | 0.25 | 1 |
| Sb1 | 0.1581(1) | 0 | 0.3345(1) | 1 |

*I* – signal intensity; *σ(I)* – standard deviation of *I*; *R1, wR2* – crystallographic R-factors; $N_{par}/N_{obs}$ – ratio between number of parameters to number of observed reflections for *I>2σ(I)*; $\rho_{min}$, $\rho_{max}$ – minimum and maximum values of residual electronic density; PTM – pressure transmitting medium;

**Dataset: 017.** CIF File: DAC02_Ne_Im3_P04.cif.

| Compound: | CoSb$_3$ | PTM: Ne |
|---|---|---|
| Pressure: | 10.7 GPa | Temperature: 293 K |
| Wavelength: | 0.2952 Å | |
| Space group: | $Im\text{-}3$, S.G. #204 | |
| Z: | 8 | |

| Lattice parameters | | | |
|---|---|---|---|
| a, Å | b, Å | c, Å | V, Å$^3$ |
| 8.73950(10) | 8.73950(10) | 8.73950(10) | |
| α, ° | β, ° | γ, ° | |
| 90 | 90 | 90 | |

| Refinement information | | | |
|---|---|---|---|
| R1, I>2σ(I) | 1.72 % | ρ$_{min}$, ρ$_{max}$, e$^-$/Å$^3$ | -1.18, 1.15 |
| wR2, I>2σ(I) | 4.18 % | | -14 ≤ H ≤ 13 |
| N$_{par}$/N$_{obs}$ | 9/337 | 1.93° ≤ θ ≤ 16.77° | -16 ≤ K ≤ 16 |
| | | | -8 ≤ L ≤ 11 |

| Structural information | | | | |
|---|---|---|---|---|
| site | x | y | z | Occ. |
| Co1 | 0.25 | 0.25 | 0.25 | 1 |
| Sb1 | 0 | 0.3344(1) | 0.1581(1) | 1 |

*I* – signal intensity; σ(I) – standard deviation of *I*; *R1, wR2* – crystallographic R-factors; $N_{par}/N_{obs}$ – ratio between number of parameters to number of observed reflections for I>2σ(I); ρ$_{min}$, ρ$_{max}$ – minimum and maximum values of residual electronic density; PTM – pressure transmitting medium;

**Dataset: 018.** CIF File: DAC02_Ne_Im3_P05.cif.

| Compound: | CoSb$_3$ | PTM: Ne |
|---|---|---|
| Pressure: | 14.1 GPa | Temperature: 293 K |
| Wavelength: | 0.2952 Å | |
| Space group: | $Im\bar{3}$, S.G. #204 | |
| Z: | 8 | |

| Lattice parameters | | | |
|---|---|---|---|
| a, Å | b, Å | c, Å | V, Å$^3$ |
| 8.67360(10) | 8.67360(10) | 8.67360(10) | |
| α, ° | β, ° | γ, ° | |
| 90 | 90 | 90 | |

| Refinement information | | | |
|---|---|---|---|
| *R1, I>2σ(I)* | 1.73 % | $\rho_{min}, \rho_{max},$ e$^-$/Å$^3$ | -1.73, 1.45 |
| *wR2, I>2σ(I)* | 4.43 % | | -14 ≤ H ≤ 13 |
| *N$_{par}$/N$_{obs}$* | 9/336 | 1.95° ≤ θ ≤ 16.9° | -16 ≤ K ≤ 16 |
| | | | -8 ≤ L ≤ 11 |

| Structural information | | | | |
|---|---|---|---|---|
| site | *x* | *y* | *z* | Occ. |
| Co1 | 0.25 | 0.25 | 0.25 | 1 |
| Sb1 | 0 | 0.3342(1) | 0.1582(1) | 1 |

*I* – signal intensity; σ(*I*) – standard deviation of *I*; *R1, wR2* – crystallographic R-factors; *N$_{par}$/N$_{obs}$* – ratio between number of parameters to number of observed reflections for *I>2σ(I)*; $\rho_{min}, \rho_{max}$ – minimum and maximum values of residual electronic density; PTM – pressure transmitting medium;

**Dataset: 019.** CIF File: DAC02_Ne_Im3_P07.cif.

| Compound: | CoSb$_3$ | PTM: Ne |
|---|---|---|
| Pressure: | 19.6 GPa | Temperature: 293 K |
| Wavelength: | 0.2952 Å | |
| Space group: | $Im\bar{3}$, S.G. #204 | |
| Z: | 8 | |

| Lattice parameters | | | |
|---|---|---|---|
| a, Å | b, Å | c, Å | V, Å$^3$ |
| 8.57530(10) | 8.57530(10) | 8.57530(10) | |
| α, ° | β, ° | γ, ° | |
| 90 | 90 | 90 | |

| Refinement information | | | |
|---|---|---|---|
| R1, I>2σ(I) | 1.27 % | $\rho_{min}, \rho_{max}$, e$^-$/Å$^3$ | -0.86, 0.68 |
| wR2, I>2σ(I) | 3.25 % | | -14 ≤ H ≤ 14 |
| N$_{par}$/N$_{obs}$ | 9/320 | 1.97°≤ θ ≤17.09° | -16 ≤ K ≤ 15 |
| | | | -8 ≤ L ≤ 11 |

| Structural information | | | | |
|---|---|---|---|---|
| site | x | y | z | Occ. |
| Co1 | 0.25 | 0.25 | 0.25 | 1 |
| Sb1 | 0 | 0.3341(1) | 0.1582(1) | 1 |

*I* – signal intensity; σ(I) – standard deviation of *I*; *R1, wR2* – crystallographic R-factors; N$_{par}$/N$_{obs}$ – ratio between number of parameters to number of observed reflections for I>2σ(I); $\rho_{min}, \rho_{max}$ – minimum and maximum values of residual electronic density; PTM – pressure transmitting medium;

**Dataset: 020.** CIF File: DAC02_Ne_Im3_P09.cif.

| Compound: | CoSb$_3$ | PTM: Ne |
| --- | --- | --- |
| Pressure: | 25.6 GPa | Temperature: 293 K |
| Wavelength: | 0.2952 Å | |
| Space group: | $Im\text{-}3$, S.G. #204 | |
| Z: | 8 | |

| Lattice parameters | | | |
| --- | --- | --- | --- |
| a, Å | b, Å | c, Å | V, Å$^3$ |
| 8.4823(2) | 8.4823(2) | 8.4823(2) | |
| α, ° | β, ° | γ, ° | |
| 90 | 90 | 90 | |

| Refinement information | | | |
| --- | --- | --- | --- |
| $R1$, $I>2\sigma(I)$ | 1.56 % | $\rho_{min}$, $\rho_{max}$, $e^-/Å^3$ | -1.22, 1.50 |
| $wR2$, $I>2\sigma(I)$ | 3.90 % | | $-14 \leq H \leq 14$ |
| $N_{par}/N_{obs}$ | 9/316 | $1.99° \leq \theta \leq 16.71°$ | $-15 \leq K \leq 15$ |
| | | | $-7 \leq L \leq 10$ |

| Structural information | | | | |
| --- | --- | --- | --- | --- |
| site | x | y | z | Occ. |
| Co1 | 0.25 | 0.25 | 0.25 | 1 |
| Sb1 | 0 | 0.3340(1) | 0.1583(1) | 1 |

*$I$ – signal intensity; $\sigma(I)$ – standard deviation of $I$; $R1$, $wR2$ – crystallographic R-factors; $N_{par}/N_{obs}$ – ratio between number of parameters to number of observed reflections for $I>2\sigma(I)$; $\rho_{min}$, $\rho_{max}$ – minimum and maximum values of residual electronic density; PTM – pressure transmitting medium;*

**Dataset: 021.** CIF File: DAC02_Ne_Im3_P10.cif.

| Compound: | CoSb$_3$ | PTM: Ne |
|---|---|---|
| Pressure: | 29.0 GPa | Temperature: 293 K |
| Wavelength: | 0.2952 Å | |
| Space group: | *I m -3*, S.G. #204 | |
| Z: | 8 | |

### Lattice parameters

| a, Å | b, Å | c, Å | V, Å$^3$ |
|---|---|---|---|
| 8.4360(2) | 8.4360(2) | 8.4360(2) | |
| α, ° | β, ° | γ, ° | |
| 90 | 90 | 90 | |

### Refinement information

| *R1, I>2σ(I)* | 2.05 % | $\rho_{min}, \rho_{max},$ e$^-$/Å$^3$ | -1.55, 1.43 |
|---|---|---|---|
| *wR2, I>2σ(I)* | 4.65 % | | -14 ≤ H ≤ 14 |
| *N$_{par}$/N$_{obs}$* | 9/311 | 2°≤ θ ≤16.81° | -15 ≤ K ≤ 15 |
| | | | -7 ≤ L ≤ 10 |

### Structural information

| site | x | y | z | Occ. |
|---|---|---|---|---|
| Co1 | 0.25 | 0.25 | 0.25 | 1 |
| Sb1 | 0 | 0.3339(1) | 0.1583(1) | 1 |

*I* – signal intensity; *σ(I)* – standard deviation of *I*; *R1, wR2* – crystallographic R-factors; *N$_{par}$/N$_{obs}$* – ratio between number of parameters to number of observed reflections for *I>2σ(I)*; $\rho_{min}, \rho_{max}$ – minimum and maximum values of residual electronic density; PTM – pressure transmitting medium;

**Dataset: 022.** CIF File: DAC02_Ne_Im3_P11.cif.

| Compound: | CoSb$_3$ | PTM: Ne |
|---|---|---|
| Pressure: | 31.2 GPa | Temperature: 293 K |
| Wavelength: | 0.2952 Å | |
| Space group: | *Im -3*, S.G. #204 | |
| Z: | 8 | |

| Lattice parameters | | | |
|---|---|---|---|
| a, Å | b, Å | c, Å | V, Å$^3$ |
| 8.4124(3) | 8.4124(3) | 8.4124(3) | |
| α, ° | β, ° | γ, ° | |
| 90 | 90 | 90 | |

| Refinement information | | | |
|---|---|---|---|
| *R1, I>2σ(I)* | 1.96 % | $\rho_{min}, \rho_{max},$ e$^-$/Å$^3$ | -1.09, 1.53 |
| *wR2, I>2σ(I)* | 4.90 % | | -14 ≤ H ≤ 14 |
| *N$_{par}$/N$_{obs}$* | 9/306 | 2.01°≤ θ ≤16.86° | -15 ≤ K ≤ 15 |
| | | | -7 ≤ L ≤ 10 |

| Structural information | | | | |
|---|---|---|---|---|
| site | *x* | *y* | *z* | Occ. |
| Co1 | 0.25 | 0.25 | 0.25 | 1 |
| Sb1 | 0 | 0.3339(1) | 0.1583(1) | 1 |

*I* – signal intensity; *σ(I)* – standard deviation of *I*; *R1, wR2* – crystallographic R-factors; *N$_{par}$/N$_{obs}$* – ratio between number of parameters to number of observed reflections for *I>2σ(I)*; $\rho_{min}, \rho_{max}$ – minimum and maximum values of residual electronic density; PTM – pressure transmitting medium;

**Dataset: 023.** CIF File: DAC02_Ne_Im3_P12.cif.

| Compound: | CoSb$_3$ | PTM: Ne |
|---|---|---|
| Pressure: | 32.0 GPa | Temperature: 293 K |
| Wavelength: | 0.2952 Å | |
| Space group: | *I m -3*, S.G. #204 | |
| Z: | 8 | |

| Lattice parameters | | | |
|---|---|---|---|
| a, Å | b, Å | c, Å | V, Å$^3$ |
| 8.4036(2) | 8.4036(2) | 8.4036(2) | |
| α, ° | β, ° | γ, ° | |
| 90 | 90 | 90 | |

| Refinement information | | | |
|---|---|---|---|
| R1, I>2σ(I) | 2.34 % | ρ$_{min}$, ρ$_{max}$, e$^-$/Å$^3$ | -2.02, 1.58 |
| wR2, I>2σ(I) | 5.28 % | | -10 ≤ H ≤ 7 |
| N$_{par}$/N$_{obs}$ | 10/299 | 2.01°≤ θ ≤16.11° | -15 ≤ K ≤ 15 |
| | | | -14 ≤ L ≤ 14 |

| Structural information | | | | |
|---|---|---|---|---|
| site | x | y | z | Occ. |
| Co1 | 0.25 | 0.25 | 0.25 | 1 |
| Sb1 | 0 | 0.1583(1) | 0.3340(1) | 0.995(1) |
| Sb2 | 0 | 0 | 0 | 0.060(9) |

*I* – signal intensity; σ(I) – standard deviation of *I*; *R1, wR2* – crystallographic R-factors; N$_{par}$/N$_{obs}$ – ratio between number of parameters to number of observed reflections for I>2σ(I); ρ$_{min}$, ρ$_{max}$ – minimum and maximum values of residual electronic density; PTM – pressure transmitting medium;

**Dataset: 024.** CIF File: DAC02_Ne_Im3_P13.cif.

| Compound: | CoSb$_3$ | PTM: Ne |
|---|---|---|
| Pressure: | 34.1 GPa | Temperature: 293 K |
| Wavelength: | 0.2952 Å | |
| Space group: | $Im\bar{3}$, S.G. #204 | |
| Z: | 8 | |

| Lattice parameters ||||
|---|---|---|---|
| a, Å | b, Å | c, Å | V, Å$^3$ |
| 8.4433(6) | 8.4433(6) | 8.4433(6) | |
| α, ° | β, ° | γ, ° | |
| 90 | 90 | 90 | |

| Refinement information ||||
|---|---|---|---|
| R1, I>2σ(I) | 6.02 % | ρ$_{min}$, ρ$_{max}$, e$^-$/Å$^3$ | -3.67, 5.19 |
| wR2, I>2σ(I) | 11.06 % | | -13 ≤ H ≤ 13 |
| N$_{par}$/N$_{obs}$ | 10/220 | 2.01° ≤ θ ≤ 15.8° | -10 ≤ K ≤ 7 |
| | | | -15 ≤ L ≤ 14 |

| Structural information |||||
|---|---|---|---|---|
| site | x | y | z | Occ. |
| Co1 | 0.25 | 0.25 | 0.25 | 1 |
| Sb1 | 0 | 0.1587(1) | 0.3346(1) | 0.932(1) |
| Sb2 | 0 | 0 | 0 | 0.813(17) |

*I* – signal intensity; *σ(I)* – standard deviation of *I*; *R1, wR2* – crystallographic R-factors; *N$_{par}$/N$_{obs}$* – ratio between number of parameters to number of observed reflections for *I>2σ(I)*; *ρ$_{min}$, ρ$_{max}$* – minimum and maximum values of residual electronic density; PTM – pressure transmitting medium;

**Dataset: 025.** CIF File: DAC02_Ne_Im3_P14.cif.

| Compound: | CoSb$_3$ | PTM: Ne |
|---|---|---|
| Pressure: | 36.8 GPa | Temperature: 293 K |
| Wavelength: | 0.2952 Å | |
| Space group: | *I m -3*, S.G. #204 | |
| Z: | 8 | |

| Lattice parameters | | | |
|---|---|---|---|
| a, Å | b, Å | c, Å | V, Å$^3$ |
| 8.401(7) | 8.401(7) | 8.401(7) | |
| α, ° | β, ° | γ, ° | |
| 90 | 90 | 90 | |

| Refinement information | | | |
|---|---|---|---|
| R1, I>2σ(I) | 5.78 % | ρ$_{min}$, ρ$_{max}$, e$^-$/Å$^3$ | -1.74, 2.83 |
| wR2, I>2σ(I) | 11.07 % | | -14 ≤ H ≤ 14 |
| N$_{par}$/N$_{obs}$ | 11/198 | 2.01°≤ θ ≤16.09° | -7 ≤ K ≤ 10 |
| | | | -13 ≤ L ≤ 13 |

| Structural information | | | | |
|---|---|---|---|---|
| site | x | y | z | Occ. |
| Co1 | 0.25 | 0.25 | 0.25 | 1 |
| Sb1 | 0.1587(1) | 0 | 0.3346(1) | 0.927(1) |
| Sb2 | 0 | 0 | 0 | 0.881(17) |

*I* – signal intensity; σ(I) – standard deviation of *I*; *R1, wR2* – crystallographic R-factors; N$_{par}$/N$_{obs}$ – ratio between number of parameters to number of observed reflections for I>2σ(I); ρ$_{min}$, ρ$_{max}$ – minimum and maximum values of residual electronic density; PTM – pressure transmitting medium;

**Dataset: 026.** CIF File: DAC02_Ne_Im3_P15.cif.

| Compound: | CoSb$_3$ | PTM: Ne |
|---|---|---|
| Pressure: | 40.8 GPa | Temperature: 293 K |
| Wavelength: | 0.2952 Å | |
| Space group: | *I m -3*, S.G. #204 | |
| Z: | 8 | |

| Lattice parameters | | | |
|---|---|---|---|
| a, Å | b, Å | c, Å | V, Å$^3$ |
| 8.3633(10) | 8.3633(10) | 8.3633(10) | |
| α, ° | β, ° | γ, ° | |
| 90 | 90 | 90 | |

| Refinement information | | | |
|---|---|---|---|
| *R1, I>2σ(I)* | 6.64 % | $\rho_{min}, \rho_{max},$ e$^-$/Å$^3$ | -4.03, 5.46 |
| *wR2, I>2σ(I)* | 12.39 % | | -10 ≤ H ≤ 7 |
| *N$_{par}$/N$_{obs}$* | 11/221 | 2.02° ≤ θ ≤ 16.17° | -14 ≤ K ≤ 15 |
| | | | -13 ≤ L ≤ 13 |

| Structural information | | | | |
|---|---|---|---|---|
| site | *x* | *y* | *z* | Occ. |
| Co1 | 0.25 | 0.25 | 0.25 | 1 |
| Sb1 | 0 | 0.1586(1) | 0.3347(1) | 0.928(1) |
| Sb2 | 0 | 0 | 0 | 0.862(17) |

*I* – signal intensity; σ(*I*) – standard deviation of *I*; *R1, wR2* – crystallographic R-factors; *N$_{par}$/N$_{obs}$* – ratio between number of parameters to number of observed reflections for *I>2σ(I)*; $\rho_{min}, \rho_{max}$ – minimum and maximum values of residual electronic density; PTM – pressure transmitting medium;

**Dataset: 027.** CIF File: DAC03_Ar_Im3_P00.cif.

| Compound: | CoSb$_3$ | PTM: Ar |
|---|---|---|
| Pressure: | 0.3 GPa | Temperature: 293 K |
| Wavelength: | 0.2952 Å | |
| Space group: | $Im\bar{3}$, S.G. #204 | |
| Z: | 8 | |

| Lattice parameters | | | |
|---|---|---|---|
| a, Å | b, Å | c, Å | V, Å$^3$ |
| 9.01810(10) | 9.01810(10) | 9.01810(10) | |
| α, ° | β, ° | γ, ° | |
| 90 | 90 | 90 | |

| Refinement information | | | |
|---|---|---|---|
| $R1, I>2\sigma(I)$ | 2.15 % | $\rho_{min}, \rho_{max}, e^-/Å^3$ | -0.74, 1.25 |
| $wR2, I>2\sigma(I)$ | 5.39 % | | $-17 \leq H \leq 17$ |
| $N_{par}/N_{obs}$ | 10/258 | $1.87° \leq \theta \leq 16.42°$ | $-17 \leq K \leq 17$ |
| | | | $-4 \leq L \leq 3$ |

| Structural information | | | | |
|---|---|---|---|---|
| site | x | y | z | Occ. |
| Co1 | 0.25 | 0.25 | 0.25 | 1 |
| Sb1 | 0 | 0.1579(1) | 0.3350(1) | 1 |

*I* – signal intensity; *σ(I)* – standard deviation of *I*; *R1, wR2* – crystallographic R-factors; $N_{par}/N_{obs}$ – ratio between number of parameters to number of observed reflections for $I>2\sigma(I)$; $\rho_{min}, \rho_{max}$ – minimum and maximum values of residual electronic density; PTM – pressure transmitting medium;

**Dataset: 028.** CIF File: DAC03_Ar_Im3_P01.cif.

| Compound: | CoSb$_3$ | PTM: Ar |
|---|---|---|
| Pressure: | 5.7 GPa | Temperature: 293 K |
| Wavelength: | 0.2952 Å | |
| Space group: | *I m -3*, S.G. #204 | |
| Z: | 8 | |

| Lattice parameters | | | |
|---|---|---|---|
| a, Å | b, Å | c, Å | V, Å$^3$ |
| 8.86620(10) | 8.86620(10) | 8.86620(10) | |
| α, ° | β, ° | γ, ° | |
| 90 | 90 | 90 | |

| Refinement information | | | |
|---|---|---|---|
| *R1, I>2σ(I)* | 2.19 % | $\rho_{min}, \rho_{max}, e^-/Å^3$ | -1.52, 3.46 |
| *wR2, I>2σ(I)* | 5.21 % | | $-17 \leq H \leq 16$ |
| *N$_{par}$/N$_{obs}$* | 10/263 | $2.33° \leq \theta \leq 16.48°$ | $-17 \leq K \leq 17$ |
| | | | $-3 \leq L \leq 5$ |

| Structural information | | | | |
|---|---|---|---|---|
| site | *x* | *y* | *z* | Occ. |
| Co1 | 0.25 | 0.25 | 0.25 | 1 |
| Sb1 | 0 | 0.1581(1) | 0.3346(1) | 1 |

*I* – signal intensity; σ(I) – standard deviation of *I*; *R1, wR2* – crystallographic R-factors; *N$_{par}$/N$_{obs}$* – ratio between number of parameters to number of observed reflections for *I>2σ(I)*; $\rho_{min}, \rho_{max}$ – minimum and maximum values of residual electronic density; PTM – pressure transmitting medium;

**Dataset: 029.** CIF File: DAC03_Ar_Im3_P02.cif.

| Compound: | CoSb$_3$ | PTM: Ar |
|---|---|---|
| Pressure: | 7.9 GPa | Temperature: 293 K |
| Wavelength: | 0.2952 Å | |
| Space group: | $Im\bar{3}$, S.G. #204 | |
| Z: | 8 | |

| Lattice parameters | | | |
|---|---|---|---|
| a, Å | b, Å | c, Å | V, Å$^3$ |
| 8.81180(10) | 8.81180(10) | 8.81180(10) | |
| α, ° | β, ° | γ, ° | |
| 90 | 90 | 90 | |

| Refinement information | | | |
|---|---|---|---|
| R1, I>2σ(I) | 2.13 % | $\rho_{min}, \rho_{max}, e^-/Å^3$ | -0.91, 1.18 |
| wR2, I>2σ(I) | 4.91 % | | -3 ≤ H ≤ 5 |
| N$_{par}$/N$_{obs}$ | 10/252 | 2.35° ≤ θ ≤ 16.57° | -16 ≤ K ≤ 17 |
| | | | -16 ≤ L ≤ 17 |

| Structural information | | | | |
|---|---|---|---|---|
| site | x | y | z | Occ. |
| Co1 | 0.25 | 0.25 | 0.25 | 1 |
| Sb1 | 0 | 0.3345(1) | 0.1581(1) | 1 |

*I* – signal intensity; *σ(I)* – standard deviation of *I*; *R1, wR2* – crystallographic R-factors; *N$_{par}$/N$_{obs}$* – ratio between number of parameters to number of observed reflections for *I>2σ(I)*; *ρ$_{min}$, ρ$_{max}$* – minimum and maximum values of residual electronic density; PTM – pressure transmitting medium;

**Dataset: 030.** CIF File: DAC03_Ar_Im3_P03.cif.

| Compound: | CoSb$_3$ | PTM: Ar |
|---|---|---|
| Pressure: | 10.0 GPa | Temperature: 293 K |
| Wavelength: | 0.2952 Å | |
| Space group: | $Im\text{-}3$, S.G. #204 | |
| Z: | 8 | |

| Lattice parameters | | | |
|---|---|---|---|
| a, Å | b, Å | c, Å | V, Å$^3$ |
| 8.76900(10) | 8.76900(10) | 8.76900(10) | |
| α, ° | β, ° | γ, ° | |
| 90 | 90 | 90 | |

| Refinement information | | | |
|---|---|---|---|
| R1, I>2σ(I) | 2.03 % | $\rho_{min}, \rho_{max},$ e$^-$/Å$^3$ | -0.88, 1.25 |
| wR2, I>2σ(I) | 4.97 % | | -16 ≤ H ≤ 16 |
| N$_{par}$/N$_{obs}$ | 10/252 | 1.93° ≤ θ ≤ 16.24° | -16 ≤ K ≤ 17 |
| | | | -5 ≤ L ≤ 3 |

| Structural information | | | | |
|---|---|---|---|---|
| site | x | y | z | Occ. |
| Co1 | 0.25 | 0.25 | 0.25 | 1 |
| Sb1 | 0 | 0.1582(1) | 0.3344(1) | 1 |

*I* – signal intensity; σ(I) – standard deviation of I; *R1, wR2* – crystallographic R-factors; N$_{par}$/N$_{obs}$ – ratio between number of parameters to number of observed reflections for I>2σ(I); $\rho_{min}, \rho_{max}$ – minimum and maximum values of residual electronic density; PTM – pressure transmitting medium;

**Dataset: 031.** CIF File: DAC03_Ar_Im3_P04.cif.

| Compound: | CoSb$_3$ | PTM: Ar |
|---|---|---|
| Pressure: | 13.8 GPa | Temperature: 293 K |
| Wavelength: | 0.2952 Å | |
| Space group: | $Im\bar{3}$, S.G. #204 | |
| Z: | 8 | |

| Lattice parameters | | | |
|---|---|---|---|
| a, Å | b, Å | c, Å | V, Å$^3$ |
| 8.6918(2) | 8.6918(2) | 8.6918(2) | |
| α, ° | β, ° | γ, ° | |
| 90 | 90 | 90 | |

| Refinement information | | | |
|---|---|---|---|
| R1, I>2σ(I) | 1.81 % | $\rho_{min}, \rho_{max}, e^-/Å^3$ | -1.18, 0.95 |
| wR2, I>2σ(I) | 4.45 % | | -16 ≤ H ≤ 16 |
| N$_{par}$/N$_{obs}$ | 10/243 | 2.38° ≤ θ ≤ 16.38° | -16 ≤ K ≤ 16 |
| | | | -5 ≤ L ≤ 3 |

| Structural information | | | | |
|---|---|---|---|---|
| site | x | y | z | Occ. |
| Co1 | 0.25 | 0.25 | 0.25 | 1 |
| Sb1 | 0 | 0.1582(1) | 0.3343(1) | 1 |

*I* – signal intensity; σ(I) – standard deviation of *I*; *R1, wR2* – crystallographic R-factors; N$_{par}$/N$_{obs}$ – ratio between number of parameters to number of observed reflections for I>2σ(I); $\rho_{min}, \rho_{max}$ – minimum and maximum values of residual electronic density; PTM – pressure transmitting medium;

**Dataset: 032.** CIF File: DAC03_Ar_Im3_P05.cif.

| Compound: | CoSb$_3$ | PTM: Ar |
|---|---|---|
| Pressure: | 17.0 GPa | Temperature: 293 K |
| Wavelength: | 0.2952 Å | |
| Space group: | $Im\bar{3}$, S.G. #204 | |
| Z: | 8 | |

| Lattice parameters | | | |
|---|---|---|---|
| a, Å | b, Å | c, Å | V, Å$^3$ |
| 8.6347(2) | 8.6347(2) | 8.6347(2) | |
| α, ° | β, ° | γ, ° | |
| 90 | 90 | 90 | |

| Refinement information | | | |
|---|---|---|---|
| R1, I>2σ(I) | 2.31 % | $\rho_{min}, \rho_{max},$ e$^-$/Å$^3$ | -1.17, 1.81 |
| wR2, I>2σ(I) | 5.45 % | | -16 ≤ H ≤ 16 |
| N$_{par}$/N$_{obs}$ | 10/236 | 1.96° ≤ θ ≤ 16.48° | -16 ≤ K ≤ 16 |
| | | | -3 ≤ L ≤ 5 |

| Structural information | | | | |
|---|---|---|---|---|
| site | x | y | z | Occ. |
| Co1 | 0.25 | 0.25 | 0.25 | 1 |
| Sb1 | 0 | 0.1582(1) | 0.3342(1) | 1 |

*I* – signal intensity; σ(I) – standard deviation of *I*; *R1, wR2* – crystallographic R-factors; N$_{par}$/N$_{obs}$ – ratio between number of parameters to number of observed reflections for I>2σ(I); $\rho_{min}, \rho_{max}$ – minimum and maximum values of residual electronic density; PTM – pressure transmitting medium;

**Dataset: 033.** CIF File: DAC03_Ar_Im3_P06.cif.

| Compound: | CoSb$_3$ | PTM: Ar |
|---|---|---|
| Pressure: | 19.3 GPa | Temperature: 293 K |
| Wavelength: | 0.2952 Å | |
| Space group: | *I m -3*, S.G. #204 | |
| Z: | 8 | |

| Lattice parameters | | | |
|---|---|---|---|
| a, Å | b, Å | c, Å | V, Å$^3$ |
| 8.5957(2) | 8.5957(2) | 8.5957(2) | |
| α, ° | β, ° | γ, ° | |
| 90 | 90 | 90 | |

| Refinement information | | | |
|---|---|---|---|
| *R1, I>2σ(I)* | 1.85 % | ρ$_{min}$, ρ$_{max}$, e$^-$/Å$^3$ | -1.07, 1.29 |
| *wR2, I>2σ(I)* | 4.72 % | | -16 ≤ H ≤ 16 |
| *N$_{par}$/N$_{obs}$* | 10/235 | 1.96° ≤ θ ≤ 16.56° | -16 ≤ K ≤ 16 |
| | | | -3 ≤ L ≤ 5 |

| Structural information | | | | |
|---|---|---|---|---|
| site | *x* | *y* | *z* | Occ. |
| Co1 | 0.25 | 0.25 | 0.25 | 1 |
| Sb1 | 0 | 0.1582(1) | 0.3342(1) | 1 |

*I* – signal intensity; *σ(I)* – standard deviation of *I*; *R1, wR2* – crystallographic R-factors; *N$_{par}$/N$_{obs}$* – ratio between number of parameters to number of observed reflections for *I>2σ(I)*; ρ$_{min}$, ρ$_{max}$ – minimum and maximum values of residual electronic density; PTM – pressure transmitting medium;

**Dataset: 034.** CIF File: DAC03_Ar_Im3_P07.cif.

| Compound: | CoSb$_3$ | PTM: Ar |
|---|---|---|
| Pressure: | 21.1 GPa | Temperature: 293 K |
| Wavelength: | 0.2952 Å | |
| Space group: | $Im\bar{3}$, S.G. #204 | |
| Z: | 8 | |

| Lattice parameters | | | |
|---|---|---|---|
| a, Å | b, Å | c, Å | V, Å$^3$ |
| 8.5737(12) | 8.5737(12) | 8.5737(12) | |
| α, ° | β, ° | γ, ° | |
| 90 | 90 | 90 | |

| Refinement information | | | |
|---|---|---|---|
| R1, I>2σ(I) | 3.40 % | $\rho_{min}, \rho_{max},$ e$^-$/Å$^3$ | -1.51, 1.88 |
| wR2, I>2σ(I) | 7.15 % | | -16 ≤ H ≤ 16 |
| N$_{par}$/N$_{obs}$ | 9/196 | 1.97°≤ θ ≤16.29° | -16 ≤ K ≤ 16 |
| | | | -3 ≤ L ≤ 5 |

| Structural information | | | | |
|---|---|---|---|---|
| site | x | y | z | Occ. |
| Co1 | 0.25 | 0.25 | 0.25 | 1 |
| Sb1 | 0 | 0.3340(1) | 0.1582(1) | 1 |

*I* – signal intensity; σ(I) – standard deviation of *I*; *R1, wR2* – crystallographic R-factors; N$_{par}$/N$_{obs}$ – ratio between number of parameters to number of observed reflections for I>2σ(I); $\rho_{min}, \rho_{max}$ – minimum and maximum values of residual electronic density; PTM – pressure transmitting medium;

**Dataset: 035.** CIF File: DAC04_Ar_Im3_P00.cif.

| Compound: | CoSb$_3$ | PTM: Ar |
|---|---|---|
| Pressure: | 28.7 GPa | Temperature: 293 K |
| Wavelength: | 0.2952 Å | |
| Space group: | $Im\text{-}3$, S.G. #204 | |
| Z: | 8 | |

| Lattice parameters | | | |
|---|---|---|---|
| a, Å | b, Å | c, Å | V, Å$^3$ |
| 8.4981(7) | 8.4981(7) | 8.4981(7) | |
| α, ° | β, ° | γ, ° | |
| 90 | 90 | 90 | |

| Refinement information | | | |
|---|---|---|---|
| $R1, I>2\sigma(I)$ | 4.30 % | $\rho_{min}, \rho_{max}, e^-/Å^3$ | -1.96, 2.43 |
| $wR2, I>2\sigma(I)$ | 9.42 % | | $-11 \leq H \leq 8$ |
| $N_{par}/N_{obs}$ | 9/220 | $1.99° \leq \theta \leq 16.46°$ | $-12 \leq K \leq 12$ |
| | | | $-13 \leq L \leq 11$ |

| Structural information | | | | |
|---|---|---|---|---|
| site | x | y | z | Occ. |
| Co1 | 0.25 | 0.25 | 0.25 | 1 |
| Sb1 | 0 | 0.3338(1) | 0.1585(1) | 1 |

*I* – signal intensity; σ(I) – standard deviation of I; *R1, wR2* – crystallographic R-factors; $N_{par}/N_{obs}$ – ratio between number of parameters to number of observed reflections for $I>2\sigma(I)$; $\rho_{min}, \rho_{max}$ – minimum and maximum values of residual electronic density; PTM – pressure transmitting medium;

**Dataset: 036.** CIF File: DAC04_Ar_Im3_P01.cif.

| Compound: | CoSb$_3$ | PTM: Ar |
|---|---|---|
| Pressure: | 30.4 GPa | Temperature: 293 K |
| Wavelength: | 0.2952 Å | |
| Space group: | *I m -3*, S.G. #204 | |
| Z: | 8 | |

| Lattice parameters | | | |
|---|---|---|---|
| a, Å | b, Å | c, Å | V, Å$^3$ |
| 8.4552(6) | 8.4552(6) | 8.4552(6) | |
| α, ° | β, ° | γ, ° | |
| 90 | 90 | 90 | |

| Refinement information | | | |
|---|---|---|---|
| *R1, I>2σ(I)* | 5.53 % | $\rho_{min}, \rho_{max}, e^-/Å^3$ | -2.53, 3.34 |
| *wR2, I>2σ(I)* | 11.10 % | | -13 ≤ H ≤ 11 |
| *N$_{par}$/N$_{obs}$* | 11/231 | 2°≤ θ ≤16.41° | -12 ≤ K ≤ 12 |
| | | | -11 ≤ L ≤ 9 |

| Structural information | | | | |
|---|---|---|---|---|
| site | *x* | *y* | *z* | Occ. |
| Co1 | 0.25 | 0.25 | 0.25 | 1 |
| Sb1 | 0 | 0.1587(1) | 0.3339(1) | 0.989(1) |
| Sb2 | 0 | 0 | 0 | 0.134(16) |

*I* – signal intensity; *σ(I)* – standard deviation of *I*; *R1, wR2* – crystallographic R-factors; *N$_{par}$/N$_{obs}$* – ratio between number of parameters to number of observed reflections for *I>2σ(I)*; $\rho_{min}, \rho_{max}$ – minimum and maximum values of residual electronic density; PTM – pressure transmitting medium;

**Dataset: 037.** CIF File: DAC04_Ar_Im3_P02.cif.

| Compound: | CoSb$_3$ | PTM: Ar |
|---|---|---|
| Pressure: | 33.9 GPa | Temperature: 293 K |
| Wavelength: | 0.2952 Å | |
| Space group: | *I m -3*, S.G. #204 | |
| Z: | 8 | |

| Lattice parameters ||||
|---|---|---|---|
| a, Å | b, Å | c, Å | V, Å$^3$ |
| 8.4529(11) | 8.4529(11) | 8.4529(11) | |
| α, ° | β, ° | γ, ° | |
| 90 | 90 | 90 | |

| Refinement information ||||
|---|---|---|---|
| *R1, I>2σ(I)* | 7.06 % | $\rho_{min}, \rho_{max}, e^-/Å^3$ | -3.58, 5.56 |
| *wR2, I>2σ(I)* | 15.06 % | | -13 ≤ H ≤ 10 |
| *N$_{par}$/N$_{obs}$* | 11/214 | 2.01° ≤ θ ≤ 15.15° | -12 ≤ K ≤ 12 |
| | | | -9 ≤ L ≤ 11 |

| Structural information |||||
|---|---|---|---|---|
| site | *x* | *y* | *z* | Occ. |
| Co1 | 0.25 | 0.25 | 0.25 | 1 |
| Sb1 | 0 | 0.1588(1) | 0.3340(1) | 0.932(2) |
| Sb2 | 0 | 0 | 0 | 0.820(18) |

*I* – signal intensity; *σ(I)* – standard deviation of *I*; *R1, wR2* – crystallographic R-factors; *N$_{par}$/N$_{obs}$* – ratio between number of parameters to number of observed reflections for *I>2σ(I)*; $\rho_{min}, \rho_{max}$ – minimum and maximum values of residual electronic density; PTM – pressure transmitting medium;

**Dataset: 038.** CIF File: DAC04_Ar_Im3_P03.cif.

| Compound: | CoSb$_3$ | PTM: Ar |
|---|---|---|
| Pressure: | 39.7 GPa | Temperature: 293 K |
| Wavelength: | 0.2952 Å | |
| Space group: | *Im -3*, S.G. #204 | |
| Z: | 8 | |

| Lattice parameters | | | |
|---|---|---|---|
| a, Å | b, Å | c, Å | V, Å$^3$ |
| 8.460(3) | 8.460(3) | 8.460(3) | |
| α, ° | β, ° | γ, ° | |
| 90 | 90 | 90 | |

| Refinement information | | | |
|---|---|---|---|
| *R1, I>2σ(I)* | 9.16 % | ρ$_{min}$, ρ$_{max}$, e$^-$/Å$^3$ | -4.95, 4.29 |
| *wR2, I>2σ(I)* | 16.25 % | | -4 ≤ H ≤ 10 |
| *N$_{par}$/N$_{obs}$* | 11/105 | 2.03° ≤ θ ≤ 16.12° | -12 ≤ K ≤ 11 |
| | | | -13 ≤ L ≤ 8 |

| Structural information | | | | |
|---|---|---|---|---|
| site | *x* | *y* | *z* | Occ. |
| Co1 | 0.25 | 0.25 | 0.25 | 1 |
| Sb1 | 0 | 0.3338(2) | 0.1586(2) | 0.938(4) |
| Sb2 | 0 | 0 | 0 | 0.750(50) |

*I* – signal intensity; σ(I) – standard deviation of *I*; *R1, wR2* – crystallographic R-factors; *N$_{par}$/N$_{obs}$* – ratio between number of parameters to number of observed reflections for *I>2σ(I)*; ρ$_{min}$, ρ$_{max}$ – minimum and maximum values of residual electronic density; PTM – pressure transmitting medium;

**Dataset: 039.** CIF File: DAC04_Ar_Im3_P04.cif.

| Compound: | CoSb$_3$ | PTM: Ar |
|---|---|---|
| Pressure: | 36.0 GPa | Temperature: 293 K |
| Wavelength: | 0.2952 Å | |
| Space group: | *Im -3*, S.G. #204 | |
| Z: | 8 | |

| Lattice parameters | | | |
|---|---|---|---|
| a, Å | b, Å | c, Å | V, Å$^3$ |
| 8.503(3) | 8.503(3) | 8.503(3) | |
| α, ° | β, ° | γ, ° | |
| 90 | 90 | 90 | |

| Refinement information | | | |
|---|---|---|---|
| R1, I>2σ(I) | 7.04 % | ρ$_{min}$, ρ$_{max}$, e$^-$/Å$^3$ | -2.96, 2.35 |
| wR2, I>2σ(I) | 14.18 % | | -9 ≤ H ≤ 11 |
| N$_{par}$/N$_{obs}$ | 11/177 | 2.02° ≤ θ ≤ 16.39° | -13 ≤ K ≤ 12 |
| | | | -9 ≤ L ≤ 12 |

| Structural information | | | | |
|---|---|---|---|---|
| site | x | y | z | Occ. |
| Co1 | 0.25 | 0.25 | 0.25 | 1 |
| Sb1 | 0 | 0.3339(2) | 0.1588(1) | 0.928(2) |
| Sb2 | 0 | 0 | 0 | 0.861(19) |

*I* – signal intensity; σ(I) – standard deviation of *I*; *R1, wR2* – crystallographic R-factors; *N$_{par}$/N$_{obs}$* – ratio between number of parameters to number of observed reflections for *I>2σ(I)*; ρ$_{min}$, ρ$_{max}$ – minimum and maximum values of residual electronic density; PTM – pressure transmitting medium;

**Dataset: 040.** CIF File: DAC04_Ar_Im3_P05.cif.

| Compound: | CoSb$_3$ | PTM: Ar |
|---|---|---|
| Pressure: | 29.8 GPa | Temperature: 293 K |
| Wavelength: | 0.2952 Å | |
| Space group: | *I m -3*, S.G. #204 | |
| Z: | 8 | |

| Lattice parameters | | | |
|---|---|---|---|
| a, Å | b, Å | c, Å | V, Å$^3$ |
| 8.546(3) | 8.546(3) | 8.546(3) | |
| α, ° | β, ° | γ, ° | |
| 90 | 90 | 90 | |

| Refinement information | | | |
|---|---|---|---|
| *R1, I>2σ(I)* | 6.90 % | $\rho_{min}, \rho_{max}, e^-/Å^3$ | -3.06, 3.41 |
| *wR2, I>2σ(I)* | 12.77 % | | -10 ≤ H ≤ 9 |
| *N$_{par}$/N$_{obs}$* | 11/138 | 3.12° ≤ θ ≤ 14.4° | -12 ≤ K ≤ 12 |
| | | | -13 ≤ L ≤ 11 |

| Structural information | | | | |
|---|---|---|---|---|
| site | *x* | *y* | *z* | Occ. |
| Co1 | 0.25 | 0.25 | 0.25 | 1 |
| Sb1 | 0 | 0.3338(2) | 0.1591(2) | 0.930(2) |
| Sb2 | 0 | 0 | 0 | 0.840(20) |

*I* – signal intensity; *σ(I)* – standard deviation of *I*; *R1, wR2* – crystallographic R-factors; *N$_{par}$/N$_{obs}$* – ratio between number of parameters to number of observed reflections for *I>2σ(I)*; *ρ$_{min}$, ρ$_{max}$* – minimum and maximum values of residual electronic density; PTM – pressure transmitting medium;

**Dataset: 041.** CIF File: DAC04_Ar_Im3_P06.cif.

| Compound: | CoSb$_3$ | PTM: Ar |
|---|---|---|
| Pressure: | 26.2 GPa | Temperature: 293 K |
| Wavelength: | 0.2952 Å | |
| Space group: | *I m -3*, S.G. #204 | |
| Z: | 8 | |

| Lattice parameters ||||
|---|---|---|---|
| a, Å | b, Å | c, Å | V, Å$^3$ |
| 8.598(4) | 8.598(4) | 8.598(4) | |
| α, ° | β, ° | γ, ° | |
| 90 | 90 | 90 | |

| Refinement information ||||
|---|---|---|---|
| *R1, I>2σ(I)* | 6.16 % | $\rho_{min}, \rho_{max,}$ e$^-$/Å$^3$ | -3.98, 2.67 |
| *wR2, I>2σ(I)* | 13.04 % |  | -9 ≤ H ≤ 10 |
| *N$_{par}$/N$_{obs}$* | 11/173 | 1.98°≤ θ ≤15.48° | -12 ≤ K ≤ 13 |
|  |  |  | -12 ≤ L ≤ 12 |

| Structural information |||||
|---|---|---|---|---|
| site | *x* | *y* | *z* | Occ. |
| Co1 | 0.25 | 0.25 | 0.25 | 1 |
| Sb1 | 0 | 0.1586(1) | 0.3344(2) | 0.928(1) |
| Sb2 | 0 | 0 | 0 | 0.870(16) |

*I* – signal intensity; *σ(I)* – standard deviation of *I*; *R1, wR2* – crystallographic R-factors; *N$_{par}$/N$_{obs}$* – ratio between number of parameters to number of observed reflections for *I>2σ(I)*; *ρ$_{min}$, ρ$_{max}$* – minimum and maximum values of residual electronic density; PTM – pressure transmitting medium;

**Dataset: 042.** CIF File: DAC04_Ar_Im3_P07.cif.

| Compound: | CoSb$_3$ | PTM: Ar |
|---|---|---|
| Pressure: | 20.7 GPa | Temperature: 293 K |
| Wavelength: | 0.2952 Å | |
| Space group: | $Im\bar{3}$, S.G. #204 | |
| Z: | 8 | |

| Lattice parameters | | | |
|---|---|---|---|
| a, Å | b, Å | c, Å | V, Å$^3$ |
| 8.685(3) | 8.685(3) | 8.685(3) | |
| α, ° | β, ° | γ, ° | |
| 90 | 90 | 90 | |

| Refinement information | | | |
|---|---|---|---|
| R1, I>2σ(I) | 6.52 % | $\rho_{min}, \rho_{max}, e^-/Å^3$ | -2.87, 3.39 |
| wR2, I>2σ(I) | 14.31 % | | -9 ≤ H ≤ 10 |
| N$_{par}$/N$_{obs}$ | 11/228 | 1.96° ≤ θ ≤ 16.54° | -11 ≤ K ≤ 13 |
| | | | -12 ≤ L ≤ 13 |

| Structural information | | | | |
|---|---|---|---|---|
| site | x | y | z | Occ. |
| Co1 | 0.25 | 0.25 | 0.25 | 1 |
| Sb1 | 0 | 0.1586(1) | 0.3341(1) | 0.930(1) |
| Sb2 | 0 | 0 | 0 | 0.844(15) |

*I* – signal intensity; σ(I) – standard deviation of *I*; *R1, wR2* – crystallographic R-factors; N$_{par}$/N$_{obs}$ – ratio between number of parameters to number of observed reflections for I>2σ(I); $\rho_{min}, \rho_{max}$ – minimum and maximum values of residual electronic density; PTM – pressure transmitting medium;

**Dataset: 043.** CIF File: DAC04_Ar_Im3_P08.cif.

| Compound: | CoSb$_3$ | PTM: Ar |
|---|---|---|
| Pressure: | 15.9 GPa | Temperature: 293 K |
| Wavelength: | 0.2952 Å | |
| Space group: | *Im -3*, S.G. #204 | |
| Z: | 8 | |

| Lattice parameters | | | |
|---|---|---|---|
| a, Å | b, Å | c, Å | V, Å$^3$ |
| 8.762(3) | 8.762(3) | 8.762(3) | |
| α, ° | β, ° | γ, ° | |
| 90 | 90 | 90 | |

| Refinement information | | | |
|---|---|---|---|
| *R1, I>2σ(I)* | 6.12 % | $\rho_{min}, \rho_{max},$ e$^-$/Å$^3$ | -3.49, 3.02 |
| *wR2, I>2σ(I)* | 5.81 % | | -12 ≤ H ≤ 12 |
| *N$_{par}$/N$_{obs}$* | 11/167 | 1.94° ≤ θ ≤ 14.97° | -13 ≤ K ≤ 11 |
| | | | -8 ≤ L ≤ 10 |

| Structural information | | | | |
|---|---|---|---|---|
| site | *x* | *y* | *z* | Occ. |
| Co1 | 0.25 | 0.25 | 0.25 | 1 |
| Sb1 | 0 | 0.3339(1) | 0.1587(1) | 0.930(1) |
| Sb2 | 0 | 0 | 0 | 0.843(15) |

*I* – signal intensity; *σ(I)* – standard deviation of *I*; *R1, wR2* – crystallographic R-factors; *N$_{par}$/N$_{obs}$* – ratio between number of parameters to number of observed reflections for *I>2σ(I)*; $\rho_{min}, \rho_{max}$ – minimum and maximum values of residual electronic density; PTM – pressure transmitting medium;

**Dataset: 044.** CIF File: DAC04_Ar_Im3_P09.cif.

| Compound: | CoSb$_3$ | PTM: Ar |
|---|---|---|
| Pressure: | 10.2 GPa | Temperature: 293 K |
| Wavelength: | 0.2952 Å | |
| Space group: | $Im\text{-}3$, S.G. #204 | |
| Z: | 8 | |

| Lattice parameters ||||
|---|---|---|---|
| a, Å | b, Å | c, Å | V, Å$^3$ |
| 8.860(2) | 8.860(2) | 8.860(2) | |
| α, ° | β, ° | γ, ° | |
| 90 | 90 | 90 | |

| Refinement information ||||
|---|---|---|---|
| $R1$, $I>2\sigma(I)$ | 5.34 % | $\rho_{min}$, $\rho_{max}$, $e^-/Å^3$ | -2.64, 2.52 |
| $wR2$, $I>2\sigma(I)$ | 12.07 % | | $-12 \leq H \leq 12$ |
| $N_{par}/N_{obs}$ | 11/253 | $1.92° \leq \theta \leq 14.18°$ | $-10 \leq K \leq 11$ |
| | | | $-13 \leq L \leq 10$ |

| Structural information |||||
|---|---|---|---|---|
| site | x | y | z | Occ. |
| Co1 | 0.25 | 0.25 | 0.25 | 1 |
| Sb1 | 0 | 0.1583(1) | 0.3340(1) | 0.935(1) |
| Sb2 | 0 | 0 | 0 | 0.782(14) |

*I* – signal intensity; *σ(I)* – standard deviation of *I*; *R1, wR2* – crystallographic R-factors; $N_{par}/N_{obs}$ – ratio between number of parameters to number of observed reflections for $I>2\sigma(I)$; $\rho_{min}$, $\rho_{max}$ – minimum and maximum values of residual electronic density; PTM – pressure transmitting medium;

**Dataset: 045.** CIF File: DAC04_Ar_Im3_P10.cif.

| Compound: | CoSb$_3$ | PTM: Ar |
|---|---|---|
| Pressure: | 4.8 GPa | Temperature: 293 K |
| Wavelength: | 0.2952 Å | |
| Space group: | $Im\bar{3}$, S.G. #204 | |
| Z: | 8 | |

| Lattice parameters | | | |
|---|---|---|---|
| a, Å | b, Å | c, Å | V, Å$^3$ |
| 8.9726(18) | 8.9726(18) | 8.9726(18) | |
| α, ° | β, ° | γ, ° | |
| 90 | 90 | 90 | |

| Refinement information | | | |
|---|---|---|---|
| R1, I>2σ(I) | 5.13 % | $\rho_{min}$, $\rho_{max}$, e$^-$/Å$^3$ | -2.11, 4.86 |
| wR2, I>2σ(I) | 11.32 % | | -14 ≤ H ≤ 14 |
| N$_{par}$/N$_{obs}$ | 11/296 | 1.89° ≤ θ ≤ 15.24° | -10 ≤ K ≤ 11 |
| | | | -10 ≤ L ≤ 12 |

| Structural information | | | | |
|---|---|---|---|---|
| site | x | y | z | Occ. |
| Co1 | 0.25 | 0.25 | 0.25 | 1 |
| Sb1 | 0 | 0.1581(1) | 0.3342(1) | 0.954(1) |
| Sb2 | 0 | 0 | 0 | 0.553(14) |

*I* – signal intensity; σ(I) – standard deviation of *I*; *R1, wR2* – crystallographic R-factors; N$_{par}$/N$_{obs}$ – ratio between number of parameters to number of observed reflections for *I>2σ(I)*; $\rho_{min}$, $\rho_{max}$ – minimum and maximum values of residual electronic density; PTM – pressure transmitting medium;

**Dataset: 046.** CIF File: DAC04_Ar_Im3_P11.cif.

| Compound: | CoSb$_3$ | | PTM: Ar |
|---|---|---|---|
| Pressure: | 0.0 GPa | | Temperature: 293 K |
| Wavelength: | 0.2955 Å | | |
| Space group: | $Im\text{-}3$, S.G. #204 | | |
| Z: | 8 | | |

| Lattice parameters ||||
|---|---|---|---|
| a, Å | b, Å | c, Å | V, Å$^3$ |
| 9.0823(4) | 9.0823(4) | 9.0823(4) | |
| α, ° | β, ° | γ, ° | |
| 90 | 90 | 90 | |

| Refinement information ||||
|---|---|---|---|
| R1, I>2σ(I) | 6.04 % | $\rho_{min}, \rho_{max}, e^-/Å^3$ | -2.98, 6.27 |
| wR2, I>2σ(I) | 13.01 % | | -15 ≤ H ≤ 13 |
| N$_{par}$/N$_{obs}$ | 11/348 | 1.86° ≤ θ ≤ 16.73° | -13 ≤ K ≤ 13 |
| | | | -9 ≤ L ≤ 11 |

| Structural information |||||
|---|---|---|---|---|
| site | x | y | z | Occ. |
| Co1 | 0.25 | 0.25 | 0.25 | 1 |
| Sb1 | 0 | 0.3354(1) | 0.1582(1) | 0.976(1) |
| Sb2 | 0 | 0 | 0 | 0.284(13) |

*I* – signal intensity; *σ(I)* – standard deviation of *I*; *R1, wR2* – crystallographic R-factors; N$_{par}$/N$_{obs}$ – ratio between number of parameters to number of observed reflections for *I>2σ(I)*; $\rho_{min}, \rho_{max}$ – minimum and maximum values of residual electronic density; PTM – pressure transmitting medium;

**Dataset: 047.** CIF File: DAC04_Ar_R3_P03.cif.

| Compound: | CoSb$_3$ | PTM: Ar |
|---|---|---|
| Pressure: | 39.7 GPa | Temperature: 293 K |
| Wavelength: | 0.2952 Å | |
| Space group: | *R -3*, S.G. #148 | |
| Z: | 12 | |

| Lattice parameters | | | |
|---|---|---|---|
| a, Å | b, Å | c, Å | V, Å$^3$ |
| 11.950(5) | 11.950(5) | 7.334(4) | |
| α, ° | β, ° | γ, ° | |
| 90 | 90 | 120 | |

| Refinement information | | | |
|---|---|---|---|
| R1, I>2σ(I) | 7.61 % | ρ$_{min}$, ρ$_{max}$, e$^-$/Å$^3$ | -2.23, 2.13 |
| wR2, I>2σ(I) | 20.63 % | | -3 ≤ H ≤ 13 |
| N$_{par}$/N$_{obs}$ | 31/189 | 2.02° ≤ θ ≤ 14.44° | -19 ≤ K ≤ 15 |
| | | | -12 ≤ L ≤ 10 |

| Structural information | | | | |
|---|---|---|---|---|
| site | x | y | z | Occ. |
| Co1 | 0.166667 | 0.333333 | 0.333333 | 1 |
| Co2 | 0.333333 | 0.666667 | 0.666667 | 1 |
| Sb1 | 0.3409(2) | 0.4958(2) | 0.5048(2) | 0.910(14) |
| Sb2 | 0.2762(2) | 0.3852(2) | 0.2174(2) | 0.939(15) |
| Sb3 | 0.333333 | 0.166667 | 0.666667 | 0.900(120) |

*I* – signal intensity; *σ(I)* – standard deviation of *I*; *R1, wR2* – crystallographic R-factors; *N$_{par}$/N$_{obs}$* – ratio between number of parameters to number of observed reflections for *I>2σ(I)*; *ρ$_{min}$, ρ$_{max}$* – minimum and maximum values of residual electronic density; PTM – pressure transmitting medium;